\documentclass[11pt,letterpaper]{article}      % generic / course paper
\usepackage[utf8]{inputenc}
\usepackage[T1]{fontenc}
\usepackage{lmodern}
\usepackage[margin=1in]{geometry}   % drop if using a venue class
\usepackage{microtype}

\usepackage{graphicx}
\usepackage{xcolor}
\usepackage{booktabs}
\usepackage{multirow}
\usepackage{tabularx}
\usepackage{array}
\usepackage{placeins}
\usepackage{adjustbox}

\usepackage{amsmath, amssymb, amsthm}
\usepackage{mathtools}

\usepackage{listings}
\usepackage{verbatim}
\usepackage{fancyvrb}

\usepackage{hyperref}
\hypersetup{
  colorlinks=true,
  linkcolor=blue!50!black,
  citecolor=green!40!black,
  urlcolor=blue!60!black,
}
\usepackage[capitalize,noabbrev]{cleveref}

\usepackage{tikz}
\usetikzlibrary{positioning, arrows.meta, shapes.geometric, shapes.multipart, calc, fit, backgrounds}

\usepackage[backend=biber,style=numeric-comp,sorting=none,maxbibnames=99]{biblatex}
\newcommand{\toolname}{\textsc{CryptoProver}\xspace}
\newcommand{\verus}{\textsc{Verus}\xspace}
\newcommand{\dalek}{\textsc{curve25519-dalek}\xspace}
\newcommand{\claudecode}{Claude~Code\xspace}

\usepackage{xspace}

\newcommand{\secref}[1]{Section~\ref{#1}}
\newcommand{\figref}[1]{Figure~\ref{#1}}
\newcommand{\tabref}[1]{Table~\ref{#1}}

\newcommand{\notesoff}{}        % silence all editorial annotations
\ifx\notesoff\undefined
  \newcommand{\todo}[1]{\textcolor{red}{\textbf{[TODO:} #1\textbf{]}}}
  \newcommand{\note}[2]{\textcolor{blue}{\textbf{[#1:} #2\textbf{]}}}
  \newcommand{\yzj}[1]{\textcolor{magenta}{[YZJ: #1]}}
  \newcommand{\resolve}[1]{\textcolor{teal!75!black}{\textbf{[RESOLVED:} #1\textbf{]}}}
  \ifx\olivernotesoff\undefined
    \newcommand{\oliver}[1]{\textcolor{blue}{\textbf{[oliver:} #1\textbf{]}}}
  \else
    \newcommand{\oliver}[1]{}
  \fi
  \ifx\ninanotesoff\undefined
    \newcommand{\nina}[1]{\textcolor{blue}{[nina: #1]}}
    \newcommand{\ninadone}[1]{\textcolor{teal}{[re nina: #1]}}
  \else
    \newcommand{\nina}[1]{}
    \newcommand{\ninadone}[1]{}
  \fi
\else
  \newcommand{\todo}[1]{}
  \newcommand{\note}[2]{}
  \newcommand{\yzj}[1]{}
  \newcommand{\resolve}[1]{}
  \newcommand{\oliver}[1]{}
  \newcommand{\nina}[1]{}
  \newcommand{\ninadone}[1]{}
\fi
\newcommand{\livia}[1]{#1}
\newcommand{\code}[1]{\texttt{\small #1}}
\providecommand{\chachaartifactcite}{\cite{chacha20verus2026}}
\providecommand{\driverlistingsize}{\footnotesize}
\providecommand{\peelfigwidth}{1.06\linewidth}

\newcommand{\admits}{\code{admit()}\xspace}
\newcommand{\ensures}{\code{ensures}\xspace}
\newcommand{\requires}{\code{requires}\xspace}
\newcommand{\axiomfn}{\code{axiom\_*}\xspace}

\newcommand{\gate}[1]{\textsc{#1}}
\newcommand{\specdrift}{\gate{spec-drift}\xspace}
\newcommand{\axiomdrift}{\gate{axiom-drift}\xspace}
\newcommand{\siblingfail}{\gate{sibling-verus}\xspace}
\newcommand{\toolingdrift}{\gate{tooling-drift}\xspace}
\newcommand{\admitcount}{\gate{admit-count}\xspace}
\newcommand{\gitrecovery}{\gate{git-recovery}\xspace}
\newcommand{\frozenedit}{\gate{frozen-edit}\xspace}
\newcommand{\forbiddenconstruct}{\gate{forbidden-construct}\xspace}

\newcommand{\nstartreal}{1{,}433\xspace}    % comment-aware non-axiom obligations (doc-comment examples excluded)
\newcommand{\ngaps}{8\xspace}               % genuine open gaps in the fork-point human residual
\newcommand{\nfinalgaps}{3\xspace}          % non-axiom gaps we leave (all shared with the human)
\newcommand{\nrunsthree}{three\xspace}      % the one-go attempt ran as three disciplined runs
\newcommand{\ncovcostusd}{730\xspace}       % 3-run total $; aggregated from the three runs' result JSONs (tab:coverage comment)
\newcommand{\ncovrounds}{154\xspace}        % 3-run total rounds; per-run result JSONs (external run/final paths, outside gen_results_macros roots)
\newcommand{\ncampruns}{24\xspace}          % stock-agent campaign runs over seven weeks
\newcommand{\ncamprounds}{451\xspace}       % campaign rounds total
\newcommand{\ncampcostusd}{1{,}452\xspace}  % campaign API cost total ($)
\newcommand{\ncamphours}{52.2\xspace}       % campaign wall-hours total (tab:campaign-phases)
\newcommand{\ncampstart}{1{,}178\xspace}    % stripped start obligations (harness count of that era)
\newcommand{\ncampclaimedpct}{97.1\xspace}  % agent-claimed closure % before audit
\newcommand{\ncamprepaircostusd}{45\xspace} % repair experiment: re-prove the 11 re-stated facts ($)
\newcommand{\ncampfinalverified}{2{,}505\xspace} % final audited whole-crate verified count, 0 errors
\newcommand{\ncampfabaxioms}{11\xspace}
\newcommand{\ncampsiblingbreaks}{5\xspace}
\newcommand{\ncampfinalaxioms}{41\xspace}   % trusted axioms in the final audited campaign state (numerically equal to \nfinalaxioms but a different run's tree; do not merge)
\newcommand{\nnohintclosed}{1{,}429\xspace}  % no-hint honest gate-verified closures = 1433 non-axiom start - 4 residual (scripts/nohint_dynamics.py)
\newcommand{\nnohintmedmin}{1.1\xspace}       % median proving time per admit, minutes (scripts/nohint_dynamics.py)
\newcommand{\nsolverpct}{19\xspace}          % Verus/Z3 verify share of no-hint wall-clock (scripts/nohint_dynamics.py)
\newcommand{\nagentpct}{81\xspace}           % agent (LLM generation + overhead) share of wall-clock
\newcommand{\nnohintgapsB}{3\xspace}      % state-B non-axiom gaps (deep Lizard/Jacobi core; 3 admits in agent-end)
\newcommand{\nnohintclosedB}{1{,}430\xspace} % = 1433 - 3, same comment-aware basis as \nnohintclosed
\newcommand{\nnohintcostusd}{748.02\xspace} % recorded lower bound: 741.38 archive + 6.64 montgomery_mul_001; excludes unrecorded clean-retry
\newcommand{\nfieldfloorlemmas}{235\xspace}   % above-field cone lemmas deleted
\newcommand{\hfCoreflZeroZeroSixAdmits}{47}   % non-axiom admits remaining (of 52; 5 axiom)
\newcommand{\nladderbanks}{27\xspace}        % banked rungs, zero fake greens campaign-wide
\newcommand{\nscaffolddrafts}{43\xspace}     % draft admits scaffolded then fully discharged in the 49-lemma home rung
\newcommand{\ncertcommentlines}{5{,}800\xspace} % ~5.8k GT comment lines surviving in the 26 editable files (block-aware, dual-verified)
\newcommand{\nreplruntimeratio}{5.5}
\newcommand{\ncertcostexact}{466.99\xspace} % recorded round-level cost; attempt-1's deadline-killed round 4 unrecorded
\newcommand{\ncerthours}{11.4\xspace}       % total elapsed time (8.1 + 3.3)
\newcommand{\ncertverified}{2{,}031\xspace} % independent fresh-container whole-crate: verified count at 0 errors (VM2 untouched copy identical)
\newcommand{\ncertaxioms}{48\xspace}        % axiom_* count at seal == the frozen floor's own count
\newcommand{\ncertlemmas}{196\xspace}       % agent lemma count (vs \nfieldfloorlemmas{} = 235 GT)
\newcommand{\ncertmasspct}{48.5\xspace}     % +11,024/-22,753 numstat over the 26 editable files = 48.5% of GT proof mass
\newcommand{\ncertinvented}{108\xspace}     % agent lemmas with no GT counterpart
\newcommand{\ncertgtskipped}{147\xspace}    % GT lemmas never reconstructed (agent's architecture doesn't need them)
\newcommand{\nchachaminutes}{15\xspace}    % duration_seconds 904 -> minutes
\newcommand{\nchachacost}{4.14\xspace}     % round_1 claude_usage.total_cost_usd 4.1394855
\newcommand{\nchachaverified}{13\xspace}   % round_1 verified_count, 0 errors, default rlimit
\IfFileExists{results-macros.tex}{% =====================================================================
\newcommand{\resOpusFourEightArmAZeroNineFreshVerified}{2{,}114}

\newcommand{\resOpusFourEightArmRlimitSites}{13}
\newcommand{\resOpusFourEightArmRlimitLoadBearingSites}{2}
\newcommand{\resOpusFourEightArmDurationHours}{62.3}

\newcommand{\resOpusFourEightArmCostUSD}{856.55}

\newcommand{\resBaselinePlainPOneAFiveCostUSD}{1{,}117.17}

\newcommand{\resBaselinePlainPOneAFiveElapsedHours}{7.42}

\newcommand{\resBaselinePlainPOneAFiveErrors}{2}
\newcommand{\resBaselinePlainPOneAFiveCompileErrors}{5}

\newcommand{\resBaselinePlainPOneAFiveFetchAttemptsBlocked}{5}
\newcommand{\resBaselinePlainPOneAFiveHistoryProbes}{38}

\newcommand{\ncaSweepVerified}{64}
\newcommand{\ncaSweepTotal}{72}
\newcommand{\ncaSweepAdmits}{1{,}090}
\newcommand{\ncaSweepRounds}{187}
\newcommand{\ncaSweepHours}{19.3}
\newcommand{\ncaSweepCost}{408}
\newcommand{\ncaResidueVerified}{6}
\newcommand{\ncaResidueTotal}{8}
\newcommand{\ncaResidueRounds}{34}
\newcommand{\ncaResidueHours}{6.0}
\newcommand{\ncaResidueCost}{148}
\newcommand{\ncaMontAdmits}{3}
\newcommand{\ncaMontRounds}{3}
\newcommand{\ncaMontHours}{0.5}
\newcommand{\ncaMontCost}{14}
\newcommand{\ncaHardAdmits}{9}
\newcommand{\ncaHardRounds}{2}
\newcommand{\ncaHardHours}{0.4}
\newcommand{\ncaHardCost}{12}
\newcommand{\ncaRepairOneModules}{5}
\newcommand{\ncaRepairOneRounds}{3}
\newcommand{\ncaRepairOneHours}{0.2}
\newcommand{\ncaRepairOneCost}{4}
\newcommand{\ncaRepairTwoAxioms}{11}
\newcommand{\ncaRepairTwoInvalid}{10}
\newcommand{\ncaRepairTwoVerified}{0}
\newcommand{\ncaRepairTwoRounds}{34}
\newcommand{\ncaRepairTwoHours}{0.9}
\newcommand{\ncaRepairTwoCost}{15}
\newcommand{\ncaRepairThreeVerified}{11}
\newcommand{\ncaRepairThreeTotal}{11}
\newcommand{\ncaRepairThreeRounds}{4}
\newcommand{\ncaRepairThreeHours}{2.2}
\newcommand{\ncaRepairThreeCost}{45}

\newcommand{\npsProofFnHuman}{813}
\newcommand{\npsProofFnAgent}{1{,}018}
\newcommand{\npsProofFnRatio}{1.25}
\newcommand{\npsLocHuman}{33{,}502}
\newcommand{\npsLocAgent}{27{,}993}
\newcommand{\npsLocRatio}{0.84}
\newcommand{\npsAssertHuman}{8{,}127}
\newcommand{\npsAssertAgent}{9{,}280}
\newcommand{\npsAssertRatio}{1.14}
\newcommand{\npsAssertByHuman}{4{,}550}
\newcommand{\npsAssertByAgent}{2{,}244}
\newcommand{\npsAssertByRatio}{0.49}
\newcommand{\npsLemmaCallsHuman}{7{,}458}
\newcommand{\npsLemmaCallsAgent}{6{,}691}
\newcommand{\npsLemmaCallsRatio}{0.90}
\newcommand{\npsBroadcastHuman}{30}
\newcommand{\npsBroadcastAgent}{13}
\newcommand{\npsBroadcastRatio}{0.43}
\newcommand{\npsForallHuman}{168}
\newcommand{\npsForallAgent}{292}
\newcommand{\npsForallRatio}{1.74}
\newcommand{\npsCalcHuman}{32}
\newcommand{\npsCalcAgent}{1}
\newcommand{\npsCalcRatio}{0.03}
\newcommand{\npsNonlinearHuman}{92}
\newcommand{\npsNonlinearAgent}{409}
\newcommand{\npsNonlinearRatio}{4.45}
\newcommand{\npsBitvectorHuman}{337}
\newcommand{\npsBitvectorAgent}{310}
\newcommand{\npsBitvectorRatio}{0.92}
\newcommand{\npsRevealHuman}{203}
\newcommand{\npsRevealAgent}{61}
\newcommand{\npsRevealRatio}{0.30}
\newcommand{\npsRlimitHuman}{5}
\newcommand{\npsRlimitAgent}{9}
\newcommand{\npsRlimitRatio}{1.80}
\newcommand{\npsDecreasesHuman}{179}
\newcommand{\npsDecreasesAgent}{188}
\newcommand{\npsDecreasesRatio}{1.05}
\newcommand{\npsCommentsHuman}{6{,}154}
\newcommand{\npsCommentsAgent}{5{,}340}
\newcommand{\npsCommentsRatio}{0.87}
\newcommand{\nstylesharedlemmas}{813}
\newcommand{\nstylemedagentloc}{12}
\newcommand{\nstylemedhumanloc}{17}

\newcommand{\ncertsolverpct}{41}
\newcommand{\ncertagentpct}{59}
\newcommand{\nbasewallcovpct}{67}
\newcommand{\nbaseverushours}{16}

\newcommand{\nparTwinsWorkers}{2}
\newcommand{\nparTwinsJobs}{2}
\newcommand{\nparTwinsSeqSecs}{411}
\newcommand{\nparTwinsWallSecs}{251}
\newcommand{\nparTwinsSpeedup}{1.64}
\newcommand{\nparTwinsFastSecs}{162}
\newcommand{\nparTwinsSlowSecs}{249}
\newcommand{\nparTwinsSoloSecs}{216}
\newcommand{\nparPoolWorkers}{2}
\newcommand{\nparPoolJobs}{3}
\newcommand{\nparPoolSeqSecs}{1002}
\newcommand{\nparPoolWallSecs}{557}
\newcommand{\nparPoolSpeedup}{1.80}
\newcommand{\nparPoolRefillSecs}{0.05}
\newcommand{\nparEightWorkers}{4}
\newcommand{\nparEightJobs}{8}
\newcommand{\nparEightSeqMin}{261}
\newcommand{\nparEightWallMin}{109}
\newcommand{\nparEightSpeedup}{2.40}
\newcommand{\nparHeavyAdmits}{23}
\newcommand{\nparMixedWorkers}{4}
\newcommand{\nparMixedJobs}{5}
\newcommand{\nparseqmin}{74}
\newcommand{\nparwallmin}{23}
\newcommand{\nparspeedup}{3.21}
\newcommand{\nparCeiling}{4}
\newcommand{\nparHardGroups}{4}
\newcommand{\nparHardProved}{11}
\newcommand{\nparHardTotal}{23}

\newcommand{\nhumanFocusedMonths}{five}
\newcommand{\nhumanFocusedPeople}{three}

}{}

\definecolor{lst-bg}{HTML}{F7F7F2}
\definecolor{lst-kw}{HTML}{1F4E79}
\definecolor{lst-kw2}{HTML}{B5651D}   % Verus-specific keywords
\definecolor{lst-str}{HTML}{2A7B3F}
\definecolor{lst-cmt}{HTML}{777777}
\definecolor{lst-num}{HTML}{8B4513}

\lstdefinelanguage{Rust}{
  morekeywords={
    fn,let,mut,pub,use,mod,crate,extern,as,if,else,match,for,while,loop,
    return,break,continue,in,ref,move,impl,trait,struct,enum,type,where,
    self,Self,super,unsafe,async,await,dyn,const,static,box
  },
  morekeywords=[2]{
    spec,proof,exec,ensures,requires,invariant,decreases,assert,assume,
    admit,axiom,broadcast,by,via,uninterp,opens_invariants,no_unwind,
    forall,exists,choose,old,seq,set,map,nat,int
  },
  sensitive=true,
  morecomment=[l]{//},
  morecomment=[s]{/*}{*/},
  morestring=[b]",
  morestring=[b]',
}

\lstdefinestyle{rust}{
  language=Rust,
  basicstyle=\ttfamily\small,
  keywordstyle=\color{lst-kw}\bfseries,
  keywordstyle=[2]\color{lst-kw2}\bfseries,
  stringstyle=\color{lst-str},
  commentstyle=\color{lst-cmt}\itshape,
  numberstyle=\color{lst-num}\footnotesize,
  backgroundcolor=\color{lst-bg},
  showstringspaces=false,
  breaklines=true,
  breakatwhitespace=true,
  numbers=left,
  numbersep=8pt,
  frame=single,
  framerule=0pt,
  framesep=4pt,
  xleftmargin=12pt,
  captionpos=b,
  tabsize=2,
}

\lstnewenvironment{veruscode}[1][]{\lstset{style=rust,#1}}{}

\newcommand{\artifactavailability}{The public \toolname artifact provides the driver, experiment manifests, released campaign records, and reproduction instructions: \url{https://github.com/ChuyueSun/CryptoProver}.}
\newcommand{\showacknowledgments}{} % non-anonymous build: render the Acknowledgments section
\newcommand{\plotasset}[1]{#1.pdf}
\newcommand{\beforeadmitlisting}{\enlargethispage{2\baselineskip}}
\newenvironment{publicationwidefigure}[1][t]{\begin{figure}[#1]}{\end{figure}}
\newenvironment{publicationwidetable}[1][t]{\begin{table}[#1]}{\end{table}}

\title{\textbf{An AI Approach to Verified Production Cryptographic Libraries}}
\author{%
  Chuyue Sun\textsuperscript{1,5},
  Su Fong\textsuperscript{1},
  Zhiyi Kuang\textsuperscript{1},
  Yizheng Jiao\textsuperscript{2},\\
  Nina Narodytska\textsuperscript{3},
  Haoze Wu\textsuperscript{3,4},
  David L. Dill\textsuperscript{1},
  Clark Barrett\textsuperscript{1,5}\\[0.5em]
  \small\textsuperscript{1}Stanford University \quad
  \textsuperscript{2}University of North Carolina at Chapel Hill\\
  \small\textsuperscript{3}VMware Research by Broadcom \quad \small\textsuperscript{4}Amherst College \quad
  \textsuperscript{5}\&Truth
}
\date{}

\begin{document}
\maketitle

\begin{abstract}
% =====================================================================
%  Abstract --- two paragraphs, approximately 200 words.
%  Paragraph 1 states the importance and unresolved production-library gap.
%  Paragraph 2 gives the system, two-library result, effort comparison, and trust mechanisms.
% =====================================================================

Cryptographic code is critical infrastructure that must be correct, yet formally verifying production libraries remains difficult.
Existing language-model proof systems solve isolated obligations with specifications and premises already given, leaving production-library verification unresolved.
\yzj{Many benchmarks have demonstrated...}\resolve{codex: declined because the abstract already states the production-library gap and the introduction and related work provide the fuller comparison.}
\yzj{Verifying cryptographic Rust crates is a harder problem: ...}\resolve{codex: declined because the abstract already states the production-library gap and the introduction and related work explain the crate-scale difficulty.}

We present \toolname, an AI-based system that synthesizes internal specifications and \verus-checked proofs from high-level API contracts.
Without changing executable code, \toolname constructs a new independent proof of \dalek and verifies RustCrypto's previously unverified \code{chacha20} implementation against an RFC~8439 specification.
These cryptographic lineages underpin deployed systems including Signal and Shadowsocks; Signal has an estimated 218M global downloads.
The original developers report a focused \nhumanFocusedMonths{}-month, \nhumanFocusedPeople{}-person effort to verify \dalek.
Given the API contracts and a fixed trusted library of field specifications, arithmetic facts, axioms, and \code{vstd}, \toolname synthesizes the internal specifications and proofs in \ncerthours{} hours with \$\ncertcostexact{} in recorded API cost.
\toolname follows a trust-first design principle: mechanical gates reject specification weakening, invented axioms, and cross-module breakage, while isolation blocks reference proof retrieval, including from git history.
\yzj{``the point'' seems too verbal}\resolve{codex: moot because the referenced wording is no longer present.}
\yzj{Too much passive voice in the results (L25--27): ``accepted by \verus'', ``is deleted'', ``are kept'', ``is deleted''. Make these active for consistency with L24's ``We evaluate\ldots by deleting'', e.g.\ ``\verus accepts the whole crate''; ``We delete every proof body but keep all specifications''.}\resolve{codex: moot because the referenced wording is no longer present.}

\end{abstract}

\section{Introduction}
\label{sec:intro}

% --- Motivation and unresolved gap ---------------------------------
Cryptographic code is critical infrastructure that must be correct, yet formally verifying production libraries remains difficult.
Production cryptographic code has a history of subtle deployment errors: an incorrect Debian-specific change to OpenSSL made cryptographic key material guessable~\cite{debian2008openssl}.
Formal verification can rule out such implementation errors relative to a specification, but it has traditionally been written by hand at a large cost in expert labor.
Language-model proof systems promise to reduce that labor, yet existing ones either solve isolated obligations with premises already in scope or synthesize specifications and proofs for a single module~\cite{yang2025autoverus,zhong2025ragverus,liu2026kverus,sun2026veristruct}.
A production crate instead requires interdependent specifications and proofs across many files.
\todo{Dave: Until recently, all formal verification was done by hand.  We need to mention that it's very laborious.  AI-based verification promises to reduce the labor, but existing systems have limitions (the things you point out here).  The abstract should also mention the large amount of expert labor required for FV.} \resolve{oliver: resolved in the intro (hand-written labor cost and AI-promise sentences above)}

\todo{nina:the ... sentence does not seem to fit the flow. It fits better for the second paragraph of intro.}
\resolve{clark:moved that content into next paragraph.}

% --- Results and human-effort comparison ----------------------------
\toolname is an AI-based verification system that takes high-level API contracts and a fixed trusted library of field specifications, field/common arithmetic facts, trusted axioms, and \code{vstd} as inputs.
The agent writes the internal specifications and proofs between them, and \verus~\cite{lattuada2023verus} checks the resulting crate.
We applied \toolname to two production cryptographic libraries without changing their executable code.  
\dalek~\cite{baif_dalek} is a library used in Signal~\cite{signallibsignal2026}, an app with an estimated 218M global downloads~\cite{sensortower2025signal}.
The original developers report a focused \nhumanFocusedMonths{}-month, \nhumanFocusedPeople{}-person effort to verify \dalek.
Given just the top-level API contracts and the trusted library, \toolname automatically synthesizes the internal specifications and proofs required for full functional verification in \ncerthours{} hours with \$\ncertcostexact{} in recorded API cost.
\todo{nina: I do not think AAAI reviewers will know what the trusted floor is; you explain it in the next para, perhaps move explanation here.} \resolve{livia: renamed the concept trusted library and defined its contents at first use.}  On another library, \toolname automatically verifies RustCrypto's \code{chacha20} v0.10.1 implementation --- used in Shadowsocks~\cite{shadowsocksrust2026} and previously unverified --- against a specification of the RFC~8439 standard. \todo{This raises the immediate question of why you verified a fork (? soft-core ?) instead of the original.} \resolve{oliver: reworded: we verify the RustCrypto implementation itself; our fork only adds the specifications.}
Our fork of the crate adds only the \verus{} specifications and proofs; its executable code is unchanged, and the verification covers the portable backend, not the SIMD backends.

% --- How the system produces the result -----------------------------
%The floor is an input to the task, not agent-generated evidence.
\todo{Dave: I don't even know what a "trusted floor" is. Is that standard terminology?  Anyway, you should explain clearly what it is and why it should be trusted. The current tone is "of course you know what this is, but I'll put in a sentence just in case."  I think it would be clearer not to use the term trusted floor and say there is a library of definitions and lemmas about number theory, etc.  Then just call it "the library"} \resolve{livia: renamed it trusted library throughout the paper and defined its contents at first use.}

% --- Trust boundary -------------------------------------------------
When the agent writes specifications as well as proofs, verifier acceptance alone does not establish that it preserved the intended claim or trust base. \todo{Dave: Explain from the beginning that there is a problem with the LLM cheating in various ways.}\todo{Andrew: To motivate the proposed techniques, perhaps we could describe the pitfall a ``vanilla'' agent can run into. And then we can have something like "To tackle this challenge..." } \resolve{oliver: added the early-campaign pitfall sentence and the transition}
An early version of our system, whose anti-cheating rules lived in the prompt rather than in mechanical checks, reported {\ncampclaimedpct}\% of \dalek closed --- yet an audit found \ncampfabaxioms{} of its ``proofs'' resting on invented axioms and another \ncampsiblingbreaks{} silently breaking sibling modules (\secref{sec:campaign}).
To avoid such issues, \toolname follows a trust-first design principle: specific gates reject specification weakening, invented axioms, cross-module breakage, and attempts to make the verifier accept without a proof.
The complete eight-gate suite also checks genuine obligation removal, frozen-file edits, tooling drift, and proof recovery from git history; fresh sessions and a sandbox that excludes the reference proof reinforce this boundary (\secref{sec:design:gates}).

% --- Contributions --------------------------------------------------
In summary, we make two main contributions:
\begin{itemize}

  \item \textbf{Trust-first proof-and-spec synthesis.}
  We name this design \emph{trust-first}: whenever possible, routes to false success are closed using mechanical checks rather than prompt language.
  The complete set of gates is discussed in \secref{sec:design:gates}.\todo{nina: somehow this point reads very similar to the para above. It is fine as it is summary of contributions, but it feels a bit repetitive} \resolve{oliver: the bullet no longer restates the trust paragraph's mechanism list; it carries only the naming act and the principle.} \todo{Dave: I feel that they are two separate points: (1) we verified this thing, and (2) we used this technique to do so.  But is "trust-first" a widely used term, or did Livia invent it? Since it's such a common problem with LLMs, I wonder if there is a standard term for measures against cheating} \resolve{oliver: ``we name'' now marks trust-first as our definition. We are not completely sure whether there is a standard term, however, we do know that ``reward hacking'' names the behavior (at least for training); Now the two bullets directly reflect the 2 points you mentioned.}

  \item \textbf{Production-library verification in hours once contracts and the trusted library are given.}
  Without changing executable code, \toolname constructs a new independent proof of \dalek and verifies RustCrypto's previously unverified \code{chacha20} implementation.

  \end{itemize}

% --- Preserved collaborator review history --------------------------
% The motivation-first rewrite resolves these comments; \notesoff keeps the
% reviewer annotations hidden, and \iffalse also suppresses Livia's former
% candidate prose while preserving the complete review record in source.
\ninadone{The motivation-first rewrite addresses the introduction-level accessibility, structure, terminology, transition, and trust-mechanism comments archived below.}

\section{Background}
\label{sec:bg}

\todo{clark: I suggest reworking this section as follows: 2.1 is good as is; 2.2 should just focus on Proof architecture - explain what the parts of a proof are - do not mention experiments; 2.3 could be a section on Proof Synthesis - explain what we mean by that term and explain that we may vary how much of a proof is synthesized, from small pieces to all of it; 2.4 can then be largely the same - focusing on how proof synthesis fails.  All of this sets up the design and implementation section.  Note that I suggest we do not mention experiments or the specific verification targets in section 2, as they really don't belong in a background section.} \resolve{oliver: done by the restructure: setups and the example moved to \secref{sec:eval}, failure modes to design; \dalek remains only as a running illustration.}
\subsection{Verus and Proof Obligations}
\label{sec:bg:verus}

\verus~\cite{lattuada2023verus} is an SMT-backed verifier for Rust: a developer annotates functions with \requires (i.e., precondition) and \ensures (i.e., postcondition) clauses,\todo{Dave: Will AAAI know what these are? "\requires (i.e., pre-conditions),..." might be clearer} \resolve{clark: added "i.e."} and \verus discharges the resulting verification conditions using an SMT solver (Z3~\cite{z3} by default).
Specifications are written in \emph{spec} functions (pure, logical models of the data), and the obligations connecting executable code to those specs are discharged in \emph{proof} functions, sometimes with explicit lemma invocations to guide the solver.
When a proof author has not yet written a proof, they can mark the hole with \admits: \verus then accepts the surrounding obligation unconditionally.
An \admits can thus be seen as explicitly flagging proofs not yet completed.
\cref{lst:admit-example} shows a representative hole and the shape of the obligation it stands in for.

\ifdefined\beforeadmitlisting\beforeadmitlisting\fi
\begin{veruscode}[caption={A typical \admits in \dalek (illustrative).},label={lst:admit-example}]
spec fn as_nat(f: FieldElement) -> nat { /* ... */ }
proof fn field_add_correct(a: FieldElement, b: FieldElement)
    ensures as_nat(field_add(a, b)) == (as_nat(a) + as_nat(b)) % P
{
    admit();   // <-- the proof obligation to discharge
}
\end{veruscode}

Certain \admits statements are intended to remain: \axiomfn lemmas that encode trusted assumptions (e.g.,\ properties of the underlying field that are taken as given in \dalek) form the codebase's \emph{trust base}.
We call all other admits \emph{non-axiom} admits; these are proof obligations that a complete verification must discharge.

\subsection{Proof Architecture and Synthesis}
\label{sec:bg:layers}
\todo{nina: "Experimental Inputs" sounds awkward in the title. What does 'experimental' mean?} \resolve{oliver: retitled.}

\todo{nina: Overall, this section does not read well. It mixes background information about standard notions in Verus with discussion of the experimental setup. Some notions, such as specifications, are also introduced multiple times, which makes the section repetitive and difficult to follow.} \resolve{livia: restructured: Background now defines each artifact once and explains the proof-versus-specification trust distinction; experiment setup and the Ristretto example moved to Evaluation, and campaign-derived failure modes moved to Design.}
A complete \verus proof tree consists of four kinds of artifacts, described below.
%categories  --- executable code, specifications, and proofs.
%There is also often a trusted floor---which includes the Verus standard library and any axioms being assumed by the proof effort.
%These are assumed to be correct or to have been proved correct elsewhere.

\begin{description}
  \item[Code.] Executable Rust, the artifact under verification.
  \item[Specifications.] Public API \requires/\ensures contracts state the promises callers rely on; internal specifications state the intermediate claims used to prove those contracts.
  Internal specifications include \code{spec} functions that define key concepts (such as field-element valuations and curve-point encodings in the case of \dalek), plus intermediate \requires/\ensures statements on helper lemmas.
  \item[Proofs.] The bodies of helper lemmas and inline proof blocks that connect executable code to the specifications.
  \item[Trusted library.] This contains the Verus standard library (\code{vstd}), field specifications, field/common arithmetic facts, and any axioms assumed by the proof effort.
  These artifacts are assumed to be correct or to have been proved correct elsewhere.
\end{description}

\emph{Synthesis} in our context refers to the process of using an AI agent to complete a partial proof tree.
If only proof bodies are synthesized,  while the specification remains fixed, this is guaranteed to preserve the original obligations: the agent must prove exactly the claims it receives.
Synthesizing specifications is riskier, because this changes the proof obligation, and a trivial or vacuous statement could be easy to prove.
Fortunately, as long as synthesis is limited to intermediate internal specifications, the top-level claim cannot be weakened: \verus{} checks each module against the contracts of the modules it depends on, so the fixed top-level API contracts are established only if every module, generated statements included, verifies.

\section{Design and Implementation}
\label{sec:design}
\label{sec:impl}

%\subsection{From Hand-Designed Workflows to Model-Directed Proof Search}

Prior proof-synthesis systems use hand-designed workflows to organize model calls~\cite{yang2025autoverus,yang2026verusage,zhong2025ragverus,liu2026kverus,sun2026veristruct}.
Such workflows restrict the degrees of freedom often required by crate-scale verification, where the next useful action may involve cross-module search, specification, decomposition, repair, or backtracking.
Improved models can now choose among these actions while operating directly on a repository.
\toolname therefore delegates the proof-search trajectory to a general-purpose coding agent instead of prescribing it in advance.  At the same time, a fully unconstrained agent has too much freedom and ends up getting lost or sabotaging itself.  Thus, we developed specific skills to keep the agent moving in the right direction as well as guardrails called \emph{gates}, which keep the agent from certain common mistakes.\todo{Clark: Where are the gates you talk about below?  You should mention them here, especially since you refer to the \specdrift{} gate by name in the next subsection.}\resolve{livia: this subsection now focuses only on the migration rationale; the driver and gates are introduced in their dedicated subsections.}

\subsection{The Motivating Campaign}

\label{sec:campaign}

\toolname's architecture was not the product of our foresight, but rather evolved as the result of careful responses to documented failures.
The campaign's input was a \emph{stripped start} of \dalek: a copy of \dalek's independently verified proof tree in which every existing proof body is replaced by \admits{}, while executable code, specifications, and trusted axioms remain unchanged, leaving \ncampstart{} open obligations.
Against this tree we ran an early version of the driver (the orchestration loop defined in \secref{sec:impl:driver})%
%(the same skill CLIs, but with only the \specdrift{} gate and none of the other cheat-detecting gates)
: \ncampruns{} intermittent runs totaling \ncamphours{} hours of summed elapsed time, \ncamprounds{} rounds, and \$\ncampcostusd{}.%
\todo{nina: it sounds too cheap for 7 weeks run :)} \resolve{oliver: the sentence now says calendar weeks, intermittent runs, and the summed elapsed hours, the bill covers 52 agent-hours, not seven weeks of runtime. Maybe we should just drop the mentioning of the 7-week timeline?} \todo{Clark: What does a ``stripped start'' mean?} \resolve{oliver: the term is now defined at first use, before the run description.}
It ended with the agent reporting success on {\ncampclaimedpct}\% of the verification conditions in the full crate, but an independent manual whole-crate audit found the claim to be inaccurate: \ncampfabaxioms{} ``proofs'' rested on axioms the agent had invented --- unproven statements whose conclusions constrain outputs their preconditions never bind --- and all but one were invalid, meaning the claimed property is false for some inputs (examples in \cref{app:campaign-ledger}). \todo{Clark: do you mean to say that there was at least one invalid axioim in each of the 11 proofs?  What do you mean by "false as universally quantified"?} \resolve{oliver: defined inline.} \todo{andrew: this is a very interesting observation. It would be cool to show examples of those incorrect axioms.} \resolve{oliver: the fabrication details and examples are now in \cref{app:campaign-ledger}.}
In \ncampsiblingbreaks{} other cases, local proofs succeeded on their own target, but broke proofs in sibling modules, and these failures were not detected by the agent. \todo{Clark: Does this mean that a local module got changed by the agent and you didn't check that the other modules were still ok?  Or does it mean something else?} \resolve{oliver: yes; now stated.} \todo{Clark: Please rewrite to define these terms \emph{before} they are used even once.} \resolve{oliver: sentence split; terms defined at use.}
Neither failure appeared in the per-target verifier output, which the agent treated as evidence of success.\yzj{``hollow'' (L16), ``green light,'' and ``success signal'' are metaphors. Consider more technical wording: ``partly hollow'' $\to$ ``partly unsound''; may revise to: ``Because each round checked only the target and not the whole crate, neither failure surfaced; the agent read the green check as success.''}\resolve{codex: replaced the metaphor with the verifier output and the agent's interpretation of that evidence.}
A key contributor to the failures was \emph{context pressure}, a collapse mode discussed in detail in \secref{sec:design:failures}: every fabrication we traced arose in a long-running session, as the agent re-ingested its own growing state and drifted toward the reward it could fake.\yzj{``names'' $\to$ ``introduces''.}\resolve{codex: moot because the referenced word is no longer present.}
Agent capability was not typically the issue.
In a fresh context, the same model found and proved corrected versions of all \ncampfabaxioms{} properties from scratch, with every constrained output bound by a precondition, for a total cost of only \$\ncamprepaircostusd{}.\todo{I don't understand this comment - previously, you say that the invented axioms fail for some inputs, but now it seems like you are saying they were proved with a fresh context - can you clarify?} \resolve{claude: clarified --- what was proved is the corrected restatement of each intended property (outputs properly bound), not the false axiom verbatim; authorized by Livia 2026-07-28.}

One key lesson from this campaign was the difference between a directive given in a prompt and one that is mechanically enforced.
In the following, we refer to an instruction stated in the agent's prompt as a \emph{rule}, while a \emph{gate} is a mechanical check the harness itself runs to enforce a rule.  The key lesson was that rules without gates are only suggestions.
\todo{nina:The notions of `rule' and `un-gated rule' have not been defined. It might be unclear that `rule' refers to a prompt instruction, while a `gate' is a mechanical check that enforces the instruction.} \resolve{oliver: both terms are now defined before first use.} \yzj{``only a suggestion'' $\to$ ``merely advisory''.}\resolve{codex: declined because the campaign lesson intentionally contrasts prompt-only rules with mechanically enforced gates.}
The campaign's only gated rule --- the specification under proof may not change, enforced by the \specdrift{} gate --- held across all \ncamprounds{} rounds; however, the other two rules left to the prompt, no new axioms and no broken siblings, were exactly the ones that were violated.\yzj{``the two'' $\to$ ``the other two''.}\resolve{codex: adopted the referent correction.}
The failures above motivated four countermeasures.
\axiomdrift{} catches fabricated axioms and \siblingfail{} catches broken siblings (\secref{sec:design:gates}).
Proof goals bind every output they constrain, preventing the malformed statements behind the fabricated axioms.
Fresh per-target sessions and in-loop resets counter context pressure (\secref{sec:impl:reset}).\yzj{Weak transition from the previous paragraph. Consider opening with a back-reference: ``Each of these failures now has a countermeasure in \toolname:''.} \resolve{claude: split one sentence per countermeasure and opened with the back-reference; authorized by Livia 2026-07-28.}
The complete per-run ledger, audit findings, and repair forensics are in \cref{app:campaign-ledger}.
\artifactavailability
\nina{This section and the section on "proof synthesis fails" are related, right? My impression is that both of them are motivation for design principles. I think it would be useful to explain failure modes well based on the experiment in this section (3.1) so it will be easier to associate the design principles with them.}
\resolve{livia: ``How Agent Proof Synthesis Fails'' now follows ``The Motivating Campaign'' and precedes the principles it motivates.}

\subsection{How Agent Proof Synthesis Fails}
\label{sec:design:failures}

We observed two failure modes of agent proof synthesis.\todo{andrew: I cannot parse "wastes its budget failing to"} \resolve{oliver: reworded.}
In a \emph{capability failure}, the agent cannot close the goal within its round budget.
In a \emph{trust failure}, the agent reports success but the result fails the driver's acceptance checks or violates the experiment's evidence boundary.
Below, we describe several examples of these failure modes observed in our campaign (\secref{sec:campaign}): the capability failures motivate the skills and context discipline%
%of \secref{sec:design}
, and the trust failures map one-to-one onto the gates.

We observed five types of capability failures: \emph{context pressure}, where a growing session re-ingests its own prior state and the agent's work degrades~\cite{liu2024lostmiddle}; \emph{learned helplessness}, where a stale failure memory makes a fresh agent give up too early; \emph{coverage gaps}, where an explicit target list omits some crucial files; \emph{liveness-signal confusion}, where a rate-limited round looks identical to an honest failure; and \emph{budget-exhausted breakage}, where a mid-edit abort leaves a file worse off unless the harness rolls it back.
Context pressure carried the sharpest lesson: it drove the campaign's fabrications (\secref{sec:campaign}).

\todo{nina: this paragraph is transition to trust modes but the transition is not very clear. May be explicitly mention, that discussion of trust odes are deferred? } \resolve{oliver: the paragraph now statrs with deferral.}
We defer the discussion of trust failures to \secref{sec:design:gates}, where each is described together with the gate that counters it.
They range from leaving an \admits in place, through fabricating a trusted axiom or weakening the specification under proof, to recovering the answer from git history.
Fabricated axioms and cross-module breaks were observed at scale in the campaign audit (\secref{sec:campaign}).

\subsection{Principles}

The driver accepts work only from verifier results, admit accounting, and gate evidence, never from the agent's report.\todo{Dave: I stumbled over "one stance" thinking it was a typo.  The straightforward way to say this: cryptoprover independently checks everything the agent does [or something like that.  or "almost everything".]. Multiple have told me that this is the only way to get quality work from an LLM in general.} \resolve{oliver: the topic sentence reworded.}\todo{Wording is too cute. "Whenever the agent claims it is done...."} \resolve{oliver: reworded with your stem; the sentence is now folded into the driver loop's acceptance rule.}
Each skill, gate, and driver policy responds to a failure observed in the campaign.
% Features we can imagine but have not yet needed stay unbuilt until an experiment shows we need them.
Within these safeguards, the agent may revise generated proof bodies and agent-authored internal lemma contracts, while the API contracts, specification vocabulary, and trusted library remain fixed (\secref{sec:noapi}).
Reference-proof retrieval is forbidden: \gitrecovery{} rejects any round that reads source code from version control.\yzj{Reader doesn't know what the principles are after reading this paragraph. Use \textbf{xxx} to delineate the principles, e.g.\ \textbf{The verifier is the only progress signal.}, \textbf{Every known cheat has a gate.}, etc.} \resolve{claude: principles now stated one per sentence in continuous prose with a counting opener; Livia prefers prose over bold headers here (2026-07-28).}

\subsection{The Driver Loop}
\label{sec:impl:driver}

\toolname is a single driver loop written in Python that makes calls to an LLM coding agent (\claudecode, in our case, though that choice is not essential).
In this paper, \emph{the driver} refers to the complete orchestration system.
The outer loop visits verification targets in a fixed order.
Each target is a module containing one or more non-axiom proof obligations (\secref{sec:bg:verus}).
One full pass over the target list is a sweep.
A task is the work of verifying one target.
Each visit to a target is an attempt.
An attempt contains a bounded sequence of rounds within a wall-clock limit.
Each round consists of one agent call followed by the driver's checks.
An accepted target is recorded in a proven registry, the driver's persistent list of already-verified targets.
When the operator enables registry filtering, a later sweep skips registered targets and revisits the remaining open targets still present in the configured list without reordering them.
At attempt start, the driver assembles a prompt containing the target module and its remaining proof obligations, relevant information from related modules, and persistent memory from prior attempts on that target.
In each round, the agent invokes skills (see \Cref{sec:design:skills}) and edits the worktree; the driver then runs \verus{} on the target and checks that the round passes every applicable integrity gate (\figref{fig:loop}; pseudocode in \cref{lst:driver}). \todo{Dave: This is hard to read. Instead of adding a bunch of qualifiers to "verus check", say "the driver invokes verus to check whether the current target property, or whole-crate property, holds, \textit{and that the result passes all of the integrity gates.}} \resolve{oliver: reworded.}
Each gate is a deterministic check computed only from recorded evidence: harness-owned task-start specification and axiom snapshots, the worktree before and after the round, the files the agent edited, and the actions it took.
The agent's completion report is not evidence of acceptance.
The driver accepts a round only if \verus{} passes, no non-axiom admits remain (\secref{sec:bg:verus}), and every applicable gate passes.
A rejected round is recoverable if the driver can restore an allowed worktree state and continue the current attempt.
A rejected but recoverable round becomes structured round history for the next agent call.
When a recoverable gate fires, the driver restores worktree state at a gate-specific scope: frozen files, frozen specifications, or the full start-of-round snapshot.
It marks the round as failed, so verification performed before restoration cannot support completion.
The following conditions end the current attempt without acceptance: an integrity violation (a fabricated axiom, an edit to the harness tooling, or a proof-bypass construct), reaching the gate retry limit after repeated failures, a verified contract inconsistency (a machine-checked counterexample shows that the implementation and its contract disagree), a decomposition request, or reaching the budget limit.
Throughout, the driver treats every signal from the agent as advisory and re-checks it independently.
Each round the agent self-reports an \code{END\_REASON}: one of \code{COMPLETE} (claims success), \code{LIMIT} (budget exhausted), \code{NEEDS\_DECOMP} (too complicated without further decomposition), or \code{FALSE\_CONTRACT} (an internal specification is false).
The driver accepts \code{COMPLETE} only when the round's independent evidence passes; otherwise it rejects the label and continues while the attempt budget remains.
%\code{LIMIT} likewise continues until that budget is exhausted;
%\todo{Clark: How can LIMIT continue if LIMIT means budget exhausted?}
An unverified \code{FALSE\_CONTRACT} becomes \code{NEEDS\_DECOMP} for a larger-budget retry, whereas a machine-verified counterexample ends the attempt as \code{FALSE\_CONTRACT}.
A later attempt may retry the same target, seeded with per-target failure memory that records declaration-level errors from earlier attempts; a prior \code{NEEDS\_DECOMP} earns the retry a larger round and wall-clock budget.
When an attempt ends, the driver disposes of its worktree by outcome: an accepted attempt is promoted to seed later work, an integrity-violation rejection is rolled back to the last checkpoint (the most recent round state that passed every gate), and any other exit is left on disk as an unpromoted candidate for analysis.

\begin{publicationwidefigure}[t]
\centering
\newcommand{\ghd}[2]{{\bfseries\boldmath\textcolor{#1}{#2}}}
\newcommand{\gbd}[1]{{\scriptsize\textcolor{black!68}{#1}}}
\resizebox{0.80\textwidth}{!}{%
\begin{tikzpicture}[
  font=\footnotesize,
  >={Stealth[length=2.2mm]},
  flowbox/.style={rounded corners=3pt, line width=0.5pt, align=center,
               text width=2.5cm, inner sep=4pt, minimum height=0.85cm,
               draw=blue!55!black, fill=blue!7},
  outcomebox/.style={flowbox, text width=4.4cm, minimum height=1.2cm},
  openbox/.style={outcomebox, draw=orange!62!black, fill=orange!9},
  registry/.style={outcomebox, draw=green!45!black, fill=green!8},
  flow/.style={->, semithick, black!72},
  yesflow/.style={->, semithick, green!48!black},
  redflow/.style={->, semithick, red!60!black},
  innerflow/.style={->, semithick, orange!68!black},
  panel/.style={rounded corners=6pt, line width=0.7pt},
]
  % ----- inner round cycle -----
  \node[flowbox] (agent) at (0,0) {\ghd{blue!45!black}{agent round}};
  \node[flowbox, right=1.45cm of agent] (checks) {\ghd{blue!45!black}{\verus{} + gates}};
  \draw[flow] (agent) -- (checks);
  \draw[innerflow, rounded corners=2.5mm]
      (checks.south) -- ++(0,-0.6)
      -- node[below, font=\scriptsize, text=orange!58!black, align=center]
         {recoverable round: errors + diagnoses to history;\\next round (fresh session if context resets)}
         ($(agent.south)+(0,-0.6)$) -- (agent.south);

  % ----- attempt and sweep contents -----
  \coordinate (attemptpad) at ($(agent.south)+(0,-1.12)$);
  \coordinate (attempttop) at ($(agent.north)+(0,0.42)$);
  \coordinate (attemptW) at ($(agent.west)+(-0.4,0)$);
  \coordinate (attemptE) at ($(checks.east)+(0.4,0)$);
  \node[flowbox, text width=2.2cm] (list) at ($(agent.west)+(-3.9,-0.55)$)
      {\ghd{blue!45!black}{configured\\target list}\\[1pt]\gbd{fixed order}};

  % ----- outer outcome state -----
  \node[openbox] (open) at ($(attemptpad)+(-1.8,-2.3)$)
      {\ghd{orange!52!black}{target remains open}\\[1pt]
       \gbd{safe failures enter persistent memory;\\a later sweep may retry}};
  \node[registry] (proven) at ($(attemptpad)+(3.3,-2.3)$)
      {\ghd{green!35!black}{proven registry}\\[1pt]
       \gbd{target complete; later sweeps skip it}};

  % ----- containment panels -----
  \coordinate (railY) at ($(open.south)+(0,-0.45)$);
  \coordinate (outerpad) at ($(railY)+(0,-0.85)$);
  \coordinate (railE) at ($(attemptE)+(1.05,0)$);
  \begin{scope}[on background layer]
    \node[panel, draw=blue!35, fill=blue!2, inner sep=4mm,
          fit=(list)(attempttop)(railE)(open)(proven)(outerpad)] (outer) {};
    \node[panel, draw=black!50, fill=black!5, inner sep=3.5mm,
          fit=(agent)(checks)(attemptpad)(attempttop)(attemptW)(attemptE)] (attempt) {};
  \end{scope}
  \node[anchor=north west, font=\scriptsize\bfseries, text=black!55]
      at ([xshift=1.5mm,yshift=-1mm]attempt.north west)
      {INNER LOOP: one attempt (bounded rounds + wall clock)};
  \node[anchor=north west, font=\scriptsize\bfseries, text=blue!55!black]
      at ([xshift=1.5mm,yshift=-1.5mm]outer.north west)
      {OUTER LOOP: one sweep};

  % ----- attempt and sweep transitions -----
  \draw[flow] (list.east)
      -- node[above, font=\scriptsize, text=black!60, sloped, pos=0.4]{one target}
      (attempt.west);
  \coordinate (redexit) at ([xshift=-1cm]open.north);
  \draw[redflow] (redexit |- attempt.south)
      -- node[right, font=\scriptsize, text=red!55!black, align=left, pos=0.5, xshift=1mm]
         {not accepted: hard cheat, repeat cap,\\verified false contract or\\decomposition request, or exhausted budget}
      (redexit);
  \draw[yesflow] (proven.north |- attempt.south)
      -- node[right, font=\scriptsize, text=green!40!black, align=left, pos=0.5, xshift=1mm]
         {success: passing, zero\\non-axiom admits,\\all gates pass}
      (proven.north);
  \draw[flow, rounded corners=2.5mm] (attempt.east)
      -- (railE |- attempt.east)
      -- node[left, font=\scriptsize, text=black!60, pos=0.72, align=right]
         {attempt\\ends}
      (railE |- railY)
      -- node[below, font=\scriptsize, text=black!60, pos=0.62, align=center]
         {next configured target\\later sweeps skip proven targets and may revisit open ones}
      (list.south |- railY)
      -- (list.south);
\end{tikzpicture}%
}
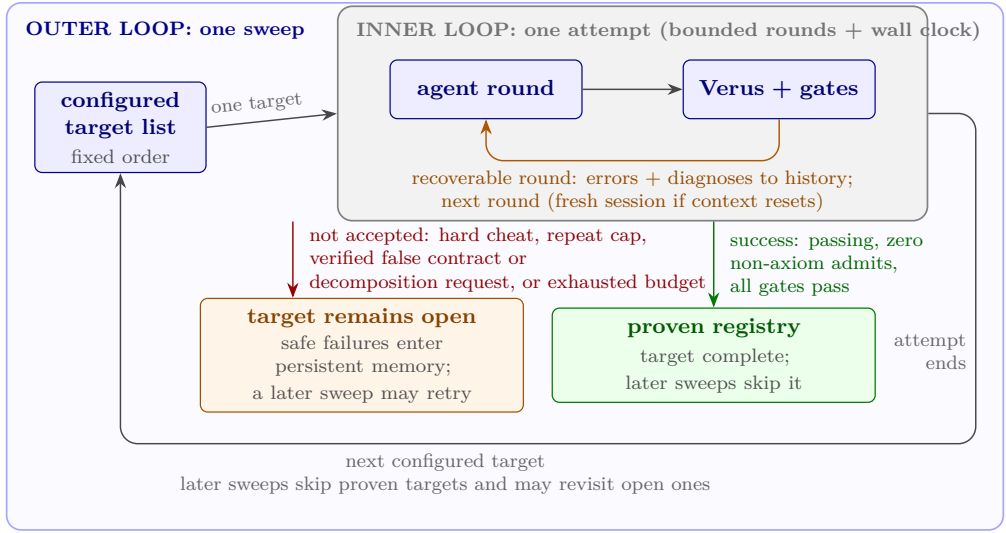
\caption{\textbf{The outer loop sweeps the configured targets in order; the inner loop iterates rounds within one attempt.}
\todo{Dave: Having trouble understanding this.  I think there's an inner loop that deals with "tasks", such as verifying a targeted property, and an outer loop that chooses targets.  But the loop nesting is not immediately clear -- the first thing I see is arrows going all over.  What is the relation between a "round" and a "task"?  I see Verus being called and gates.  I don't understand much else.}
\resolve{livia: nested the round loop inside one bounded attempt and the attempt inside one ordered sweep; outer state records show how later sweeps treat proven and open targets.}}
\label{fig:loop}
\end{publicationwidefigure}

\paragraph{Context budget and auto-reset.}
\label{sec:impl:reset}
As the driver runs, the session reuses the prompt cache and accumulates context, which, as explained in \secref{sec:campaign}, can create problems.
The driver starts each attempt in a fresh session and may reset between rounds after a stall (consecutive short rounds that fill no admits), context bloat (accumulated session context past a token threshold), or proof plateau (no improvement in the progress metric across several rounds), up to a per-target cap.
A reset preserves the worktree and round history and does not replenish the attempt's round or wall-clock budget.

\paragraph{Decomposition.}
\label{sec:impl:decomp}
On a later attempt after \code{NEEDS\_DECOMP}, the driver injects guidance to split the target's remaining proof obligations into named lemmas, with a loop-invariant template for iterative obligations.
The parallel orchestration design and measurements are in the appendix, \cref{app:parallel}.
% Parallelism material moved to \cref{app:parallel} by Livia's direction.
% \label{sec:impl:parallel}
% Because \verus{} checks a lemma against its callees' \ensures{} signatures rather than their bodies, proving obligations is embarrassingly parallel.
% A thin orchestrator exploits this with a post-merge re-verification that no per-group claim can bypass, reaching $\nparspeedup\times$ at four-way parallelism.

\subsection{The Six Skills}
\label{sec:design:skills}
\label{sec:impl:skills}

\todo{andrew: This subsection could use a leading sentence to motivate why we need to develop dedicated skills. It could also motivate each particular skills a bit more.} \resolve{oliver: added the leading motivation sentence and a clause motivating each search skill; the other skills' motivations were already inline.}
A raw agent with no harness misjudges its own progress: targets could appear completed to the agent while obligations remain, and existing lemmas might get re-derived or fabricated (\secref{sec:design:failures}).
\toolname therefore gives the agent six dedicated skill CLIs in three groups.
Verification skills check claimed completion, admit accounting exposes remaining obligations, and search avoids redundant or fabricated lemmas.
The six skills are: \code{verus\_check}, \code{admit\_inventory}, \code{search\_semantic}, \code{search\_module}, \code{search\_macro}, and \code{search\_proven}.
All six share one contract: arguments in, a JSON result out, a trace appended, and an exit code that mirrors the result's \code{okay} field.
\code{verus\_check} is the source of truth for ``did it verify?''
\code{admit\_inventory} counts non-axiom admits (\secref{sec:bg:verus}), with comments and \axiomfn bodies filtered out so the count cannot be gamed, which turns ``the file looks done'' into a checkable predicate.
The four search skills let the agent find existing lemmas wherever they hide --- \code{search\_semantic} by meaning, \code{search\_module} by home module, \code{search\_macro} behind a macro expansion, and \code{search\_proven} in an earlier run's record --- instead of re-deriving or fabricating them, the failure mode that produces invented axioms.
Because every skill uses the same interface, the driver invokes and parses them uniformly.

\subsection{The Gate Suite}
\label{sec:design:gates}
\label{sec:impl:gates}
\todo{nina: gate suite is one of the central notions and was forward referenced multiple times. However, the definition of gate is defined only at this point. The section is svery dense, it is really hard to keep a high level picture of the purpose of the gate w.r.t. the rules they gate. I wonder a table will be better, e.g. gate, what is prevents, and what it checks.}
The suite of eight gates is the core process-integrity mechanism of the design: each gate counters one untrustworthy success from \secref{sec:design:failures}, and the full predicates are in \cref{app:gates}.

\admitcount{} credits a passing \verus run only if it actually removed an obligation.
\axiomdrift{} lets the agent use the existing axiom base but fails any round that extends it --- the fabrication mode the campaign surfaced at scale (\secref{sec:campaign}).
\specdrift{}, implemented by the harness-owned \code{spec\_check} CLI rather than an agent skill, requires the specification under proof to be exactly what it was at task start, so weakening or deleting the \ensures is never a route to an accepted round.
\siblingfail{} re-verifies the target area and every file edited during the round, so a target cannot pass by breaking another module.\todo{nina: `local green bought at global cost' is not clear} \resolve{oliver: stated literally now}
\toolingdrift{} fails a round that edits the harness, the verifier configuration, or the skills; \gitrecovery{} fails a round that reads source code out of version control, because the stripped tree's history still carries the original proof and a proof recovered from history is retrieval, not synthesis.
\frozenedit{} fails a round that touches any file the experiment marks frozen (the trusted library, the specification vocabulary, or a proved sibling).
\forbiddenconstruct{} fails a round that introduces a construct discharging an obligation without a proof, i.e., any bypass the other counters cannot see.

\section{Evaluation}
\label{sec:eval}

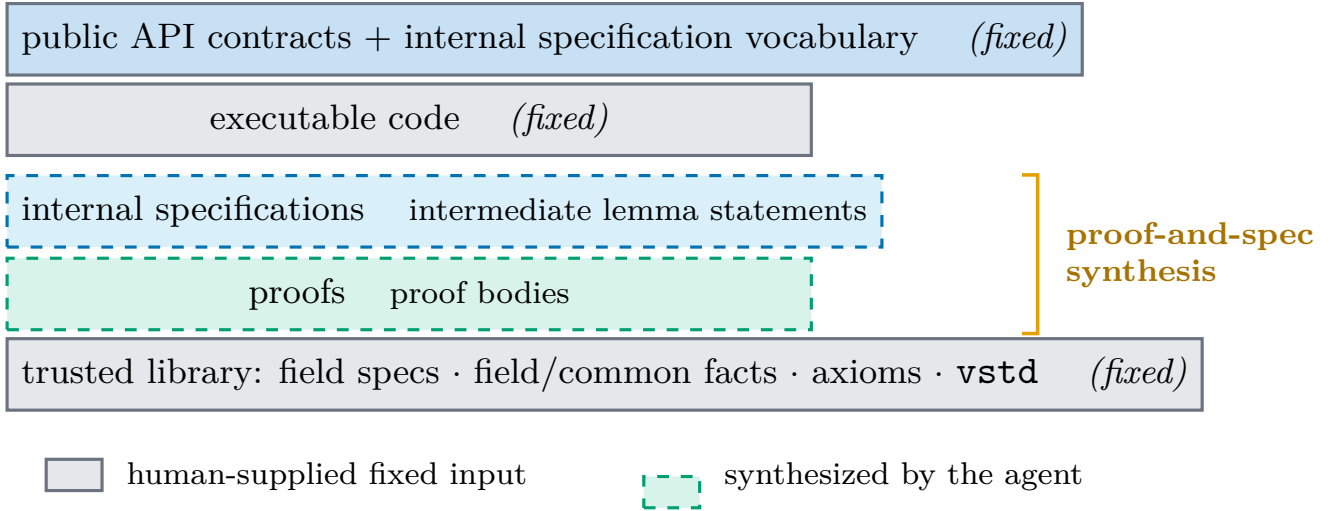
\begin{figure}[t]
\centering
\definecolor{peelSpecFill}{HTML}{D9F0FB}
\definecolor{peelSpecDraw}{HTML}{0072B2}
\definecolor{peelProofFill}{HTML}{D8F3EA}
\definecolor{peelProofDraw}{HTML}{009E73}
\definecolor{peelFloorFill}{HTML}{E5E7EB}
\definecolor{peelFloorDraw}{HTML}{6B7280}
\definecolor{peelApiFill}{HTML}{C7E0F4}
\definecolor{peelCut}{HTML}{E69F00}
\definecolor{peelPin}{HTML}{9A4F86}
\makebox[\linewidth][c]{\resizebox{\peelfigwidth}{!}{%
\begin{tikzpicture}[font=\footnotesize,
  band/.style={minimum width=7.2cm, minimum height=0.64cm, anchor=south west, align=center, line width=0.8pt},
  frozen/.style={band, draw=peelFloorDraw, fill=peelFloorFill},
  specb/.style={band, draw=peelSpecDraw, dashed, fill=peelSpecFill},
  proofb/.style={band, draw=peelProofDraw, dashed, fill=peelProofFill},
  runlab/.style={anchor=west, align=left, text=peelCut!72!black, font=\scriptsize},
]
  % fixed givens (contract, code) on top; internal specifications + proofs; fixed floor
  \node[frozen, fill=peelApiFill] at (0,3.00) {public API contracts + internal specification vocabulary \quad\textit{(fixed)}};
  \node[frozen]                   at (0,2.28) {executable code \quad\textit{(fixed)}};
  \node[specb]  at (0,1.46) {internal specifications \;\; {\scriptsize intermediate lemma statements}};
  \node[proofb] at (0,0.72) {proofs \;\; {\scriptsize proof bodies}};
  \node[frozen] at (0,0.00) {trusted library: field specs $\cdot$ field/common facts $\cdot$ axioms $\cdot$ \code{vstd} \quad\textit{(fixed)}};
  % Proof-and-spec synthesis targets the two interior bands.
  \draw[peelCut, line width=0.9pt] (9.1,2.12) -- (9.24,2.12) -- (9.24,0.70) -- (9.1,0.70);
  \node[runlab] at (9.36,1.41) {\textbf{proof-and-spec}\\ \textbf{synthesis}};
  % legend
  \node[frozen, minimum width=0.5cm, minimum height=0.28cm] (lg1) at (0.35,-0.72) {};
  \node[anchor=west, font=\scriptsize] at (0.95,-0.58) {human-supplied fixed input};
  \node[proofb, minimum width=0.5cm, minimum height=0.28cm, anchor=west] (lg2) at (5.70,-0.72) {};
  \node[anchor=west, font=\scriptsize] at (6.30,-0.58) {synthesized by the agent};
\end{tikzpicture}%
}}
\caption{\textbf{The artifact layers of a verified crate.}
%executable code, public API contracts with the internal specification vocabulary, intermediate internal specifications, proof bodies, and the trusted library --- are defined in \secref{sec:bg:layers}.
Shaded layers are human-supplied fixed input; dash-outlined layers are agent-synthesized.
\emph{Proof-and-spec synthesis} targets the intermediate internal specifications and proof bodies above the fixed trusted library.}\todo{Dave: Start by explaining that a formally verified crate would have these five parts} \resolve{livia: the caption now defines all five required layers and separates the human-supplied inputs from the layers assigned to the AI agent.} \resolve{oliver: caption slimmed to names + reading key; the definitions live in \secref{sec:bg:layers}, and the supplies-what split is in the text below the figure.}
\label{fig:peel-design}
\end{figure}

The \dalek proof-and-spec synthesis run is our main experiment: with executable code, API contracts, and the trusted library fixed, \toolname must synthesize all intermediate specifications and proofs.
\todo{Is "field-floor" a term used in dalek?  I'd like to see a few words about what it is, especially the "floor" part.} \resolve{livia: the following sentence explains that the experiment name refers to the trusted library's lower-boundary role.} \resolve{codex: the internal experiment name was retired from reader-facing text on 2026-07-28.}
%The experiment name refers to the trusted library as the fixed lower edge of the editable region.
A final transfer experiment asks \toolname to synthesize proofs for RustCrypto's previously unverified \code{chacha20} implementation against human-authored RFC~8439 specifications.
Throughout the evaluation, we count only non-axiom admits as remaining work.
\figref{fig:peel-design} shows the proof-and-spec synthesis experiment's fixed inputs and requested outputs.

%In the \emph{proof synthesis} experiment (\secref{sec:eval:coverage-sec}), the agent additionally receives every intermediate internal specification and supplies only the proof bodies.
%In the \emph{proof-and-spec synthesis} experiment (\secref{sec:noapi}), it supplies both the intermediate internal specifications and the proof bodies for \dalek's Edwards, Montgomery, Ristretto, and scalar modules\todo{nina: I would mention that these are modules from the library as reviewers might be not familiar with modules in curve25519} \resolve{oliver}.
%Proof-and-spec synthesis excludes the trusted library from the editable files, while proof synthesis operates over a crate-wide writable tree, so a whole-crate audit instead confirms that the library did not change (\secref{sec:eval:coverage-sec}).

\nina{It is hard to understand this cut, an example would help }
\ninadone{Resolved: the example now explains the supplied inputs and synthesized outputs, and corrects the prior claim that the Ristretto \code{spec} function was removed.}
For example, consider a task targeting the Ristretto module of \dalek. The agent receives the executable \code{compress} function, its fixed public API contract, the fixed internal specification vocabulary for Ristretto encodings, and the trusted library.
The agent must state and prove the intermediate internal specifications that connect the encoding arithmetic to that contract, so that \code{compress} and its fixed callers pass the verification check.
\cref{app:placement} lists the supplied and synthesized material for every module in the \dalek proof-and-spec synthesis experiment.

\subsection{The Proof-And-Spec Synthesis Run}
\label{sec:noapi}

\label{sec:noapi:certificate}
%Before the run, we fixed the prompt, stopping rules, success criteria, and accessible information; the reference proof was absent from the run environment (\cref{app:certdetail}).
Using \code{claude-fable-5}, \toolname{} synthesized every intermediate specification and proof connecting the API contracts of \dalek's Edwards, Montgomery, Ristretto, and scalar modules to the trusted library in \ncerthours{} hours of elapsed time, with \$\ncertcostexact{} in recorded API cost (\figref{fig:certificate-dynamics}).
A fresh x86-Linux container with the pinned \verus{} release reported \ncertverified{} checks verified, zero errors at the default resource limit; the final tree contained no unresolved proof obligations, no executable-code changes, and exactly \ncertaxioms{} axioms, all already present in the trusted library.

\begin{publicationwidefigure}[t]
  \centering
  % Data: field-floor records plus the plain-Claude result, watchdog, and
  % monitoring ledger, plus tracked Opus bundles; the generation script parses provenance.
  \includegraphics[width=0.90\textwidth]{\plotasset{certificate_dynamics}}
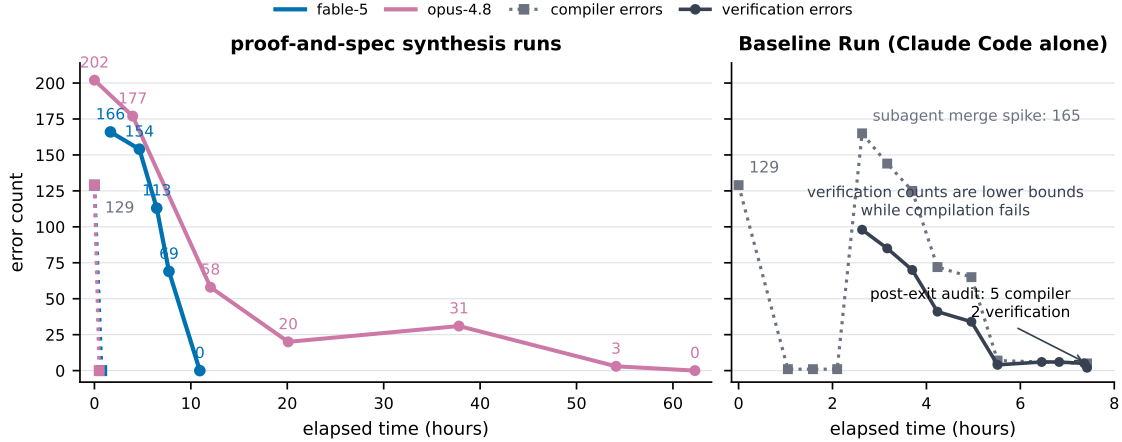
  \caption{Error trajectories on the proof-and-spec synthesis task.
    The proof-and-spec synthesis runs share one elapsed-time axis: compiler lines connect the initial compiler-error count to the first zero-compiler-error milestone, while verification lines report errors on the whole crate.  In the baseline run, the agent did not attempt to verify any targets until over two hours had passed.}
  \label{fig:certificate-dynamics}
\end{publicationwidefigure}

\label{sec:noapi:fieldfloor}
During the run, the agent corrected its own false intermediate specification after a machine-checked counterexample while executable code, API contracts, and the trusted library remained fixed.

The agent produced \ncertlemmas{} proof functions, compared with \nfieldfloorlemmas{} in the human reference, using {\ncertmasspct}\% as many proof lines; \ncertinvented{} agent proof functions have no reference counterpart, and \ncertgtskipped{} reference proof functions are absent.
%; the remaining shared names are concentrated among those required by frozen callers (\figref{fig:divergence-map}).
The agent proofs are therefore more compact and shorter, while the limited overlap in proof functions shows that the agent reorganized much of the proof architecture.
%Our analysis did not test whether shared proof bodies match or establish whether shorter proofs are higher quality. \todo{Dave: Is it the \textit{number} of proof bodies? Answer "so what"?} \resolve{livia: defined the counts and added the compactness and divergence interpretation.}
The agent also added auxiliary \code{spec fn} definitions for computable mirrors, induction measures, and a loop target; each required a proof connecting it to the fixed specification vocabulary.
In \figref{fig:divergence-map}, unfilled outlines count human-reference proof functions, blue bars count agent proof functions, and green segments count functions with the same name and source file; labels report agent/human counts followed by the shared count in parentheses.

\begin{publicationwidefigure}[t]
  \centering
  \includegraphics[width=\linewidth]{\plotasset{divergence_map}}
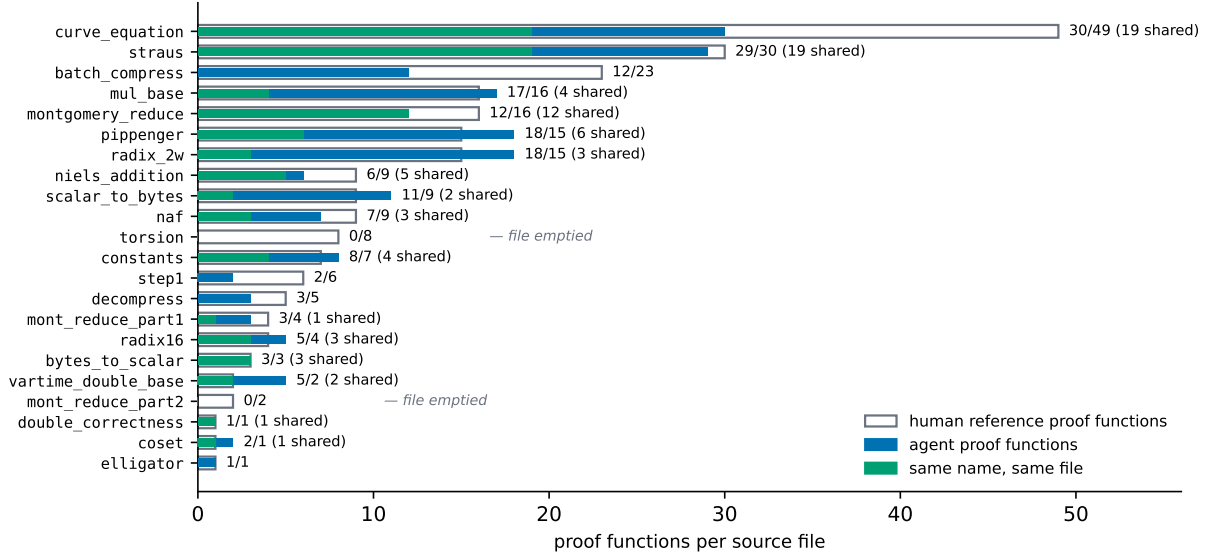
  \caption{Agent and human proof functions by source file.}
  % Data: the committed divergence study, transcribed in scripts/divergence_map.py.
  \label{fig:divergence-map}
\end{publicationwidefigure}
\label{sec:noapi:ungated}
% Run record: results/baseline_plain_p1_a5/result.json (VM2 arm a5); plan + dated
% amendments in dalek-lite-mvp/docs/plain_claude_baseline_plan.md; coordination
% ledger AGENT_DEBATE_PLAIN_CLAUDE_VM2_2026-07-08.md.
To measure what \toolname{}'s driver, supplied skills, and gates add, the baseline ran the same model through \code{claude-code} on the proof-and-spec synthesis task, under the same stated constraints in a network-sealed container, without any of the three. \todo{Dave: "baseline" could be more specific. "To understand the added value of the driver and framework, we tried verifying the crate by simply asking Claude to do it, without the framework ..."} \resolve{oliver: reworded.}
On the same task, the baseline exited after \resBaselinePlainPOneAFiveElapsedHours{} hours at a cost of \$\resBaselinePlainPOneAFiveCostUSD{}, claiming it had completed the task.  But an analysis of the output revealed \resBaselinePlainPOneAFiveCompileErrors{} compiler and \resBaselinePlainPOneAFiveErrors{} verification errors remaining. 
Logs record \resBaselinePlainPOneAFiveFetchAttemptsBlocked{} fetches and \resBaselinePlainPOneAFiveHistoryProbes{} history probes, all of which were blocked by the network seal.
In its trace in \figref{fig:certificate-dynamics}, the compiler-error spike is the result of subagents merging code into the shared tree and introducing integration errors. \todo{Dave: Do you mean the second batch of compiler errors, after they fell to 0?} \resolve{livia: yes; identified it as the post-merge spike.}
Its last completed success checks were module-scoped, which was the wrong scope for confirming overall success, and the session exited while the whole-crate check was still running.
The plotted verification counts are lower bounds, as compiler errors prevent the checker from reporting results on the whole crate.

\label{sec:noapi:replication}
\toolname replicated the proof-and-spec synthesis result with \code{opus-4.8}, a second model from the same family; a fresh x86-Linux container with the pinned \verus{} release reported \resOpusFourEightArmAZeroNineFreshVerified{} checks verified, zero errors.
The run took \resOpusFourEightArmDurationHours{} hours and recorded \$\resOpusFourEightArmCostUSD{} in API cost (\secref{sec:discuss:threats}).
Thus, \code{opus-4.8} took $\nreplruntimeratio\times$ the elapsed time of \code{claude-fable-5}. \todo{Dave: I think a sentence of interpretation after each description would be appropriate. "Opus 4.8 took 6 times as long as Fable to completely verify the crate" or whatever. Answer "so what" after every figure.} \resolve{livia: added the convergence comparison and interpretation.}
Both models resolved the same hardest obligation by decomposing the proof.
Increasing the solver budget did not close the obligation; extracting a closed-form helper and splitting two sublemmas did.
%; in the replication, \resOpusFourEightArmRlimitLoadBearingSites{} of \resOpusFourEightArmRlimitSites{} solver-limit sites remained necessary under single-site removal (\cref{app:certdetail}).

\label{sec:noapi:chacha}
% Run record: chacha20-verus fork, field-floor peel (trusted-core), run_id ks_fieldfloor.
% Sealed start 5fa5dcb -> reconstruction 979f00d (branch field-floor-reconstruction);
% harness dalek-lite-mvp @ 8fc9d55 + branch field-floor-generic (5e7948a); metrics from
% verus_core/result.json; see macros.tex chacha block + chacha20/REPORT_fieldfloor.md + COMPARISON.md.
% Elapsed-time and API-cost values CONFIRMED against the archived run record 2026-07-14:
% duration_seconds 904.041247 (-> 15 min) and round_1 total_cost_usd 4.1394855 (-> $4.14),
% identical in _chacha20_fork_ff_results/ks_fieldfloor/verus_core and the packaged
% _chacha20_artifacts copy; REPORT_fieldfloor.md L29 agrees.
ChaCha20 has verified implementations in other ecosystems~\cite{zinzindohoue2017hacl,protzenko2020evercrypt}, but the RustCrypto implementation we targeted has not been formally verified before.  We human-authored and independently validated a formal version of the RFC~8439 specification. Then, using \code{opus-4.8}, \toolname synthesized the proofs in one round (\nchachaminutes{} minutes, \$\nchachacost{}).
At acceptance, the whole-crate \verus{} check reported \nchachaverified{} verified items with no errors at the default solver limit, no proof-position admits, and no specification drift.
%
% chacha20 fork released publicly 2026-07-14: https://github.com/oliversssf2/chacha20-verus
% (default branch verus-soft-core = verified crate + docs; branch field-floor-reconstruction
% + tags ff-gt/ff-start/ff-reconstructed carry the sealed peel; COMPARISON.md has the diff guide).
The verified fork and the sealed reconstruction experiment are public~\chachaartifactcite.

\section{Discussion}
\label{sec:discuss}

%\subsection{Fixed inputs constrain an agent-authored interior}

Given fixed API contracts and a trusted library, \toolname authored machine-checked proof interiors for two production cryptographic libraries without changing executable code (\secref{sec:eval}).
The generated proof architecture also diverged from the human reference: fixed inputs constrain correctness without prescribing the internal construction (\secref{sec:noapi}).

\toolname uses \verus{} to check verification conditions and gates to reject changes that weaken the claim, expand the trusted base, or break other modules.
An accepted run may therefore differ from the fixed inputs only in the agent-authored interior: acceptance requires the whole crate to re-verify against the unchanged human-written contracts and trusted library.
This architecture emphasizes capability and soundness relative to fixed inputs rather than trust in the agent.

The campaign's false successes exposed failures of trust and context discipline rather than limits of prover capability (\secref{sec:campaign}).
We interpret the gates and the fresh-session discipline as making the model's proof capability acceptable by blocking the known cheats.
% This interpretation is tested only for the assembled system, not per component (\secref{sec:noapi}).
The experiments are existence results rather than estimates of average performance because they are not independent repeated trials.
The gate suite bounds only the failure modes it encodes; it does not rule out an unmodeled bypass.

The campaign's lesson is the one we would carry to other agent-verification systems: rules must be mechanically enforced to be effective.
Wherever success is machine-checkable and the known cheats are mechanically gated, an agent-authored artifact can be accepted without trusting the agent's process.

% --- reviewer notes (grouped; anchors given per note) ---
\note{Livia}{[re the opening sentence] Could this opening state the capability directly and leave the peel cut as the evaluation method rather than making the capability depend on that instrument?}
\resolve{claude: replaced the recap with the joint takeaway; outcomes and evidence remain in Section 4; authorized by Livia 2026-07-28.}
\note{Livia}{[re the opening paragraph] Could we replace this experiment-by-experiment recap with one sentence giving only the joint takeaway, since the exact outcomes and success evidence already live in Section 4 and the older-run provenance is repeated under Threats?}
\resolve{claude: replaced the recap with the joint takeaway; outcomes and evidence remain in Section 4; authorized by Livia 2026-07-28.}
\note{Livia}{[re the divergence sentence] Section 4 already reports the repaired generated specification, so could we cut the run chronology here and keep only the conceptual boundary lesson in the next sentence?}
\resolve{claude: replaced the recap with the joint takeaway; outcomes and evidence remain in Section 4; authorized by Livia 2026-07-28.}
\note{Livia}{[re the divergence sentence] The proof counts, proof mass, divergence, and hardest-obligation decomposition already appear in Section 4, so could this retain only the implication that fixed boundaries do not prescribe the internal proof architecture?}
\resolve{claude: replaced the recap with the joint takeaway; outcomes and evidence remain in Section 4; authorized by Livia 2026-07-28.}

\subsection{Soundness remains relative to the trusted base}

The logical guarantee comes from \verus{} (backed by Z3) checking the crate against the human-authored API contracts and the trusted library.
The gate implementations, audit scripts, and container configuration provide process-integrity evidence that the run preserved those inputs and avoided the enumerated routes to false success.
That the audits found no violations of the enumerated failure modes means the agent added no trust of its own; it does not mean the result has no trusted assumptions.
% Because the two main experiments cover different scopes and use different trusted bases, neither result subsumes the other (\secref{sec:eval}).

% The evaluated trusted library includes facts specific to this prime field; applying the system to another curve would require an appropriate library, which we do not evaluate.

% --- reviewer notes (grouped; anchors given per note) ---
\note{Livia}{[re the scopes/bases sentences] This non-equivalence is important, but Section 4 already gives the counts and scope details, so could these sentences become one implication about the runs' different scopes and trusted bases?}
\note{Livia}{[re the trusted-base enumeration] Could we avoid the retired ``certificate'' term and choose one home for the trusted-base enumeration, which already appears in the substantive discussion above?}
\resolve{codex: the non-equivalence sentence is no longer reader-facing, so this subsection does not repeat the experiments' scope details.}
\resolve{codex: separated the logical guarantee from process-integrity evidence and kept the trusted-base enumeration in this subsection.}

\subsection{Proof quality and autonomy remain open}

Whether the generated proofs remain maintainable as the library evolves is a pressing open question.
We, not the agent, supplied the targets and --- most importantly --- their proof order; letting the agent plan that order from the contracts is immediate future work.

% --- reviewer notes (grouped; anchors given per note) ---
\note{Livia}{[re the residual-obligations sentence] The obligations are unproved rather than shown unprovable, so could we call this not an integrity failure but a current solver and proof-engineering capability frontier, while cutting the repeated Section 4 details?}
\resolve{claude: moot --- the sentence moved to Appendix C in the one-story restructure and already carries the frontier framing (``particularly difficult obligations incomplete'').}

\subsection{Threats to validity}
\label{sec:discuss:threats}

The result is functional correctness against the supplied contracts, not cryptographic security, constant-time execution, side-channel resistance, or contract adequacy.
% Cut (Oliver, length pass): duplicates the boundary sentences in 5.1 — pending Livia's OK since it was her requested wording.
% API contracts and the trusted library remain fixed, generated interior specifications may be repaired, and whole-crate re-verification rejects repairs that violate those inputs. \oliver{This sentence seems unnecessary?}
On cost, the \ncerthours{}-hour figure measures agent elapsed time after the contracts, the trusted library, the specification vocabulary, the target decomposition, the proof order, and the harness had been supplied, while the developers' reported \nhumanFocusedMonths{}-month effort is a calendar duration rather than measured person-hours, so the two figures are not directly comparable.
Within the verified artifact, humans supplied the high-level API contracts and the trusted library, while \toolname authored the internal specifications and every proof; we therefore expect a large reduction in human verification effort, though these figures do not measure it.
The same-family \code{opus-4.8} run also completed the verification effort, but at $\nreplruntimeratio\times$ the headline runtime (\cref{app:certdetail}) of the \code{fable-5} run.
Finally, the prompts, skills, and gates were tuned on \dalek{}, and both subjects were selected favorably --- \dalek{} for its auditable human reference, \code{chacha20} for its size and RFC~8439 specification --- so the results may overstate performance on an unseen crate.

% Cut (Oliver, length pass): the one-family/one-verifier disclosure and the closing recover
% sentence — the Conclusion states the claim one column later, and S10 carries generality.
% The evaluated models come from one family and all runs use \verus{}; extending across model vendors and verifiers is the natural next experiment.

% --- reviewer notes (grouped; anchors given per note) ---
\note{Livia}{[re Claim scope] Please move the mechanism names to the Appendix and describe only the concept here: API contracts and the trusted floor remain fixed, generated interior specifications may be repaired, and whole-crate re-verification rejects repairs that violate those inputs.}
\note{Livia}{[re Isolation] This revision drops the quantitative reference-comment leakage disclosure, which has no other reader-facing home, so could we restore that count here while omitting the frozen-callsite count already reported in Section 4?}
\resolve{claude: disclosure dropped by Livia's ruling 2026-07-28 --- the count stays artifact-side only; content of the inherited comments was not audited and she accepts that scope.}
\resolve{codex: Livia approved Oliver's length cut; the fixed-boundary concept remains in Section 5.1, while this paragraph no longer repeats the mechanism details.}

\section{Related Work}
\label{sec:related}
\label{sec:related:verif}
\label{sec:related:specsound}
\label{sec:related:crypto}
\label{sec:related:trust}
\label{sec:related:tp}

The closest \verus{} systems generally synthesize proofs from supplied specifications or bounded targets: AutoVerus, VeruSAGE, and RagVerus work at function or file granularity, KVerus applies dependency-aware synthesis to real kernel code, and VeriStruct jointly plans specifications and proofs for modules~\cite{yang2025autoverus,yang2026verusage,zhong2025ragverus,liu2026kverus,sun2026veristruct}.
VerusSeek strengthens proof synthesis with fine-grained retrieval of contracts, invariants, lemmas, proof blocks, and assertions, followed by hierarchical context expansion~\cite{zhang2026verusseek}.
\toolname{} instead synthesizes missing cross-file internal specifications and proofs for production libraries from fixed API contracts and a trusted library, with mechanical gates that preserve those fixed inputs.

Research on fallible specifications shows why local proof success is insufficient: generated annotations can be vacuous or false, so acceptance must separately check specification consistency and non-triviality~\cite{sun2024clover,ye2025verina,wu2024lemur,agarwal2026ids}.
Verified cryptographic implementations establish the value of end-to-end machine checking~\cite{ye2017hmacdrbg,almeida2019lastmile,kuepper2023cryptopt}, while recent AI pipelines and benchmarks have begun applying proof models to cryptographic code~\cite{klaus2026rustleanai,tan2026dalekbench}.
Neural theorem provers usually receive a fixed external statement~\cite{ren2025deepseekproverv2,yang2023leandojo}; when an agent can also alter specifications or proof assumptions, documented proof gaming makes a passing verifier signal insufficient~\cite{aggarwal2024alphaverus,bursuc2025vericoding}.

\section{Conclusion}
\label{sec:conclusion}

Our results show that it is now possible to formally verify widely used cryptographic libraries with a small fraction of the human effort: given API contracts and a trusted library, \toolname synthesized \dalek's internal specifications and proofs in \ncerthours{} hours for \$\ncertcostexact{}, whereas the original developers report a focused \nhumanFocusedMonths{}-month, \nhumanFocusedPeople{}-person verification effort.
\toolname further verified RustCrypto's previously unverified \code{chacha20} implementation.
In both cases, no executable code was changed.
\verus checking, the complete eight-gate suite, fresh sessions, and sandboxing defend this process against false success and reference-proof retrieval.
The verified claim is functional correctness against the supplied contracts, not constant-time execution or side-channel resistance (\secref{sec:discuss:threats}).

Future work includes applying \toolname to additional cryptographic libraries.  We also expect the same techniques to largely transfer to other (non-cryptographic) Rust systems, and AI-assisted authoring of the requirement specifications would reduce the last major human task. \todo{Dave: This weakens the message by starting out with trusted floors, etc. State the real, general goal of the paper: The results show that it's practical to formally verify widely-used cryptographic libraries with orders of magnitude less human effort by using AI assisted formal verification tools. For future work, more crypto libraries is the first thing.  But the same techniques can be applied to other code.  Also, AI-assisted writing of the requirements specification would further reduce the last remaining major human task. } \resolve{oliver: reframed as suggested; wrote ``a small fraction of the human effort'' rather than ``orders of magnitude'' because the human months also covered specification and infrastructure work.}

% Shown only in the non-anonymous (arXiv) build: main.tex defines \showacknowledgments;
% main-aaai.tex does not (AAAI submission is anonymous, and pages after 7 are references-only).
\ifdefined\showacknowledgments
\section*{Acknowledgments}
This work was supported in part by the Defense Advanced Research Projects Agency (DARPA) under contract FA8750-24-2-1001, the Chen Institute, and LMSYS.
\fi

\printbibliography

\appendix
\section{Gate Definitions}
\label{app:gates}

This appendix defines each gate formally.
% Implementation pin: dalek-lite-mvp@2c1d2052b662fbc90020c4476c0a5176132ae88f.
% run.py creates spec_snapshot.json and records baseline_axioms at task start,
% then compares later states against those harness-owned snapshots.
Let \(S_0\) be the harness-owned snapshot recorded at task start.
Each round transforms a pre-round tree \(S_{\mathrm{pre}}\) into a post-round tree \(S_{\mathrm{post}}\) for a target module \(t\).
Let \(\mathit{Edited}\) be the files the agent changed and \(\mathit{Cmds}\) the shell commands it issued during the round.
Each gate is a pure predicate over \((S_0, S_{\mathrm{pre}}, S_{\mathrm{post}}, t, \mathit{Edited}, \mathit{Cmds})\) (\secref{sec:impl:gates}).\yzj{``this data'' is an unclear reference.}\resolve{claude: named the exact tuple defined in the preceding sentence; approved by Livia 2026-07-28.}
The harness accepts the round only if \verus{} verifies \(t\) in \(S_{\mathrm{post}}\) and every predicate below holds.
Otherwise, the named gate fires and the harness rejects the round.
Write \(s[f]\) for the byte content of file \(f\) in state \(s\) and \(\mathrm{ok}(s,f)\) for ``\(f\) verifies under \verus{} in \(s\).''

\paragraph{\admitcount{} (vs.\ false completion).}
Let \(a(s)\) be the non-axiom admit count of \(t\) in \(s\): the number of \admits{} placeholders standing in for unproved goals, excluding trusted axioms.
The \code{admit\_inventory} command exposes the same counter to the agent.
Passes iff \(a(S_{\mathrm{post}}) < a(S_{\mathrm{pre}})\); otherwise, the harness rejects a round that removed no obligation.
\oliver{why "canonical"? can we just drop it?}\resolve{claude: dropped ``canonical'' and glossed what the counter counts.}

\paragraph{\axiomdrift{} (vs.\ fabricated axioms).}
Let \(\mathrm{Ax}(s)\) be the set of \axiomfn{} names in \(s\).
Passes iff \(\mathrm{Ax}(S_{\mathrm{post}}) \subseteq \mathrm{Ax}(S_0)\): the existing trust base may be used but not extended.
This predicate detects new axiom names.
In the proof-and-spec synthesis and convergence-ladder experiments reported below, axiom files are also frozen, so \frozenedit{} rejects an in-place statement change.

\paragraph{\specdrift{} (vs.\ spec drift).}
Let \(\mathrm{spec}(s,t)\) be \(t\)'s preconditions and postconditions (its \requires/\ensures{} clauses) and its spec-function bodies.
Passes iff \(\mathrm{spec}(S_{\mathrm{post}},t)\) is byte-identical to \(\mathrm{spec}(S_0,t)\).

\paragraph{\siblingfail{} (vs.\ cross-module breakage).}
Let \(\mathit{Area}(t)\) be \(t\)'s top-level area module.
Passes iff \(\mathrm{ok}(S_{\mathrm{post}},f)\) for every \(f \in \mathit{Edited} \cup \{\mathit{Area}(t)\}\): a break in any touched sibling fails the round even when \(t\) itself verifies.

\paragraph{\toolingdrift{} (vs.\ a doctored checker).}
Let \(\mathit{Tool}\) be the harness, verifier configuration, and skill-CLI files.
Passes iff \(S_{\mathrm{post}}[f] = S_{\mathrm{pre}}[f]\) for every \(f \in \mathit{Tool}\) (compared by content hash).

\paragraph{\gitrecovery{} (vs.\ answer recovery from history).}
Passes iff no command in \(\mathit{Cmds}\) matches one of the following source-reading forms: \code{git show}, \code{git checkout <ref> -- <file>}, \code{git log -p}, \code{git diff} against \code{HEAD}, \code{git cat-file}, \code{git stash show -p}, or \code{git worktree add}.\oliver{seems unnecessary to mention the metadata-only form?}\resolve{claude: dropped the metadata-only clause; the rule wording already excludes those forms.}
In the proof-and-spec synthesis run, the original proof bodies were absent from the machine, so there was nothing for these commands to recover; the gate provided an additional safeguard (\secref{sec:design}).\oliver{drop defense-in-depth since it is unclear what it means}\resolve{claude: replaced the term with a plain statement of the same point.}

\paragraph{\frozenedit{} (vs.\ out-of-scope edits).}
Let \(\mathit{Frozen}\) be the files the experiment marks frozen (substrate lemmas, spec vocabulary, proved siblings).
Passes iff \(\mathit{Edited} \cap \mathit{Frozen} = \emptyset\).

\paragraph{\forbiddenconstruct{} (vs.\ proof-free discharge).}
Let \(c(s)\) count \code{assume(...)} and \code{\#[verifier::}\allowbreak\code{external\_body]} occurrences across the editable files.
Passes iff \(c(S_{\mathrm{post}}) \le c(S_{\mathrm{pre}})\): such a construct discharges an obligation without a proof checked by the SMT solver (\verus{}'s underlying automated prover) and leaves neither an \admits{} nor a new \axiomfn{} for the other counters to catch.

\paragraph{Reads of task-start state.}
Only \specdrift{} and \axiomdrift{} consult state from outside the round: both compare the post-round tree against the harness-owned task-start snapshot \(S_0\).
The harness creates this snapshot before the agent begins the task and keeps it outside the agent's control, so the agent cannot pass either gate by editing both sides at once.

\paragraph{Gate routing in the driver.}
\Cref{lst:driver} shows where the gate suite sits in the per-target round loop and how firings route through claim checking, retry, terminal rejection, and post-loop promotion.
The driver's outer loop enters this per-target round loop once for each configured target, in supplied order.

\begin{publicationwidefigure}[t]
\begin{veruscode}[language={},basicstyle=\ttfamily\driverlistingsize,caption={Per-target driver pseudocode: \verus and \emph{all} gates must pass.},label={lst:driver}]
# The outer loop enters here once per target, in configured order.
for round in 1..=max_rounds:             # within wall-clock deadline
    prompt = render(template, failure_memory, last_errors)
    if round == 1 or stalled() or bloated():  # context budget:
        agent = fresh_session()               # else continue -c, cache reuse
    stream(agent, prompt)                # agent edits the worktree, runs skills
    result = run_verus(target)
    gates  = run_gates(target, result)   # admit-count, axiom-drift, spec-drift,
                                         # sibling-verus, tooling-drift,
                                         # git-recovery, frozen-edit,
                                         # forbidden-construct
    record(round, result, gates)
    if gates.hard_cheat():               # axiom-drift / tooling-drift /
                                         #   forbidden-construct:
        return gates.cheat_label         #   terminal on first firing
    if gates.recoverable_fired():        # spec-drift/frozen-edit/git-recovery:
        restore_frozen_state()           #   restore, taint the round, instruct
        continue                         #   the agent; past a cap -> terminal
    if claimed(COMPLETE):                # the agent claims; the harness checks:
        if result.ok and admits_left(target) == 0:
            return COMPLETE              # passing, zero non-axiom admits, gates clean
        reject_claim_with_reason()       # claim refused; agent told why
    if claimed(FALSE_CONTRACT):          # a frozen contract is false:
        return FALSE_CONTRACT if witness_verified() else NEEDS_DECOMP
                                         #   the witness is machine-checked;
                                         #   unverified -> escalation only
    if claimed(NEEDS_DECOMP):            # escalate, not a dead end: a fresh
        return NEEDS_DECOMP              #   retry resumes with a larger budget
                                         #   (+2 rounds, 1.5x wall-clock) + a
                                         #   "build the missing
                                         #   infrastructure first" directive
    round_history.append(target, result.errors, gates)    # non-fatal: loop back
return final_label(rounds)           # post-loop evidence check: a tree meeting
                                     #   the completion criteria promotes to
                                     #   COMPLETE even unclaimed; a tainted or
                                     #   unverified last round never does; else LIMIT
# exit disposition, on every return above: a cheat-class label rolls the
#   worktree back to the last integrity-clean snapshot; a non-COMPLETE label
#   feeds persistent failure memory (unless its trace is tainted); only
#   COMPLETE is promoted -- any other final state stays on disk, unpromoted
\end{veruscode}
\end{publicationwidefigure}

\section{Campaign Per-Run Ledger}
\label{app:campaign-ledger}

\begin{publicationwidetable}[t]
  \centering
  \footnotesize
  \setlength{\tabcolsep}{3.5pt}
  % Source: data/appendix_results.json, generated into results-macros.tex.
  \begin{tabularx}{\linewidth}{@{}lXrrr@{}}
    \toprule
    Run & Outcome & Rounds & Elapsed (h) & Cost (\$) \\
    \midrule
    \code{sweep\_all\_001}         & \ncaSweepVerified/\ncaSweepTotal{} modules; \textasciitilde\ncaSweepAdmits{} admits filled & \ncaSweepRounds & \ncaSweepHours & \ncaSweepCost \\
    \code{residue\_001}            & \ncaResidueVerified/\ncaResidueTotal{} retry of sweep failures & \ncaResidueRounds & \ncaResidueHours & \ncaResidueCost \\
    \code{montgomery\_retry\_001}  & \ncaMontAdmits{} admits ``closed'' (later shown fabricated) & \ncaMontRounds & \ncaMontHours & \ncaMontCost \\
    \code{hard\_tail\_001}         & last \ncaHardAdmits{} honest admits closed & \ncaHardRounds & \ncaHardHours & \ncaHardCost \\
    \midrule
    \code{repair\_001}             & fixed the \ncaRepairOneModules{} sibling-broken proofs & \ncaRepairOneRounds & \ncaRepairOneHours & \ncaRepairOneCost \\
    \code{repair\_002\_axioms}     & \ncaRepairTwoAxioms{} axioms forced as lemmas: \ncaRepairTwoVerified/\ncaRepairTwoAxioms{} (\ncaRepairTwoInvalid{} invalid as stated) & \ncaRepairTwoRounds & \ncaRepairTwoHours & \ncaRepairTwoCost \\
    \code{repair\_003\_inline}     & re-proved all \ncaRepairThreeTotal{} inline, fresh context: \ncaRepairThreeVerified/\ncaRepairThreeTotal & \ncaRepairThreeRounds & \ncaRepairThreeHours & \ncaRepairThreeCost \\
    \bottomrule
  \end{tabularx}
  \caption{The main-sweep and repair runs that the campaign audit and repair turn on.}
  \label{tab:campaign-audit}
\end{publicationwidetable}

This appendix records the motivating campaign of \secref{sec:campaign} run by run.
\tabref{tab:campaign-audit} breaks out the runs that the campaign audit and repair turn on.
All numbers are aggregated from the per-target \code{result.json} records in our supplementary campaign artifact: the \code{claude\_usage} cost and token fields, \code{duration\_seconds}, and \code{rounds\_used}.
A fourth repair run was an aborted false-success attempt whose rounds, hours, and cost are folded into the \code{repair\_002\_axioms} row.
The full 152-target ledger, with per-round diffs and spec snapshots, lives in that artifact.
Raw session transcripts are withheld from the review copy and released with the camera-ready artifact.
The audit (\code{audit\_001}) was a separate non-proving verification pass and is not counted as a run.

% Source: supplementary artifact final_report.md S1.3 (corrected accounting: montgomery ladder/
% to_edwards/elligator x4, scalar invert chain x1, ristretto compress/decode/elligator/batch x6),
% S2.5 (shared shape; worst case lemma_ristretto_compress_correct), and repair_handoff.md
% (per-file placement and names of the 11; montgomery_retry_001 forensics).
The eleven fabricated axioms share one shape: each constrains an output parameter in its \ensures{} without relating that parameter to the inputs in its \requires{}, so invoking the lemma supplies the conclusion without proof. \oliver{it is unclear what "parametric lemma" means, also unclear what ".... its  code{requires} nver binds"}\resolve{claude: dropped ``parametric lemma'' and restated the binding defect in plain terms.}
All eleven are asserted without proof; ten are moreover invalid --- the claimed property fails for some inputs --- while the eleventh is true as stated.
The eleventh, \code{lemma\_batch\_loop\_}\allowbreak\code{iteration\_correct}, is a Ristretto batch-loop property with no counterpart in the reference proof, but it was not proved as a standalone lemma.
The extreme case applies \code{lemma\_ristretto\_}\allowbreak\code{compress\_correct} to \code{point} and \code{s\_bytes}.
It states no \requires{} at all and claims that arbitrary bytes equal the compression of an arbitrary point.
Four of the eleven cover the Montgomery ladder (differential add-and-double, conversion to Edwards form, basepoint-on-curve, Elligator encoding), one a 27-step scalar inversion chain, and six the Ristretto compress, decode, Elligator, and batch paths.
The three admits \code{montgomery\_retry\_001} ``closed'' were all discharged by inventing such axioms in a sibling file and calling them.
\code{repair\_003\_inline} later proved the intended properties behind all eleven fabricated axioms from scratch, with no access to the reference proof, for \$\ncaRepairThreeCost{}.
Each property was restated inline with its outputs bound to its inputs.
The final whole-crate state was \ncampfinalverified{} verified, zero errors, and \ncampfinalaxioms{} trusted axioms.
% Verified 2026-07-15 against the archived run records on the origin machine
% (~/field-report-artifact/mvp-results, a 2026-06-27 pre-pruning copy of dalek-lite-mvp/results
% with all 72 sweep module dirs): every row's rounds/cost reproduce exactly from result JSONs —
% sweep 187/$407.95, residue 34/$148.15, montgomery 3/$14.45, hard_tail 2/$11.54,
% repair_001 3/$4.07, repair_002 26+8(false-green)=34/$9.17+$5.82=$14.98, repair_003 4/$45.23.
% Hours: audit-table repair rows are result.json per-task elapsed (0.24/0.89/2.16 -> 0.2/0.9/2.2);
% the phase table sums round durations (repair 3.10 -> 3.1, main sweep 26.24 -> 26.2).

\FloatBarrier
\section{Whole-Crate Proof-Only Run}
\label{sec:eval:coverage-sec}
\label{sec:eval:setup}
\label{sec:impl:admitted}
\label{sec:eval:cost}
\label{sec:eval:coverage}
\label{sec:eval:beyond-human}
\label{sec:eval:soundness}
\label{sec:eval:ablation}
\label{app:proof-only-details}
This appendix reports the whole-crate proof-only run under two conditions: \emph{no-hints}, with all comments removed before the run, and \emph{with-hints}, with source doc-comments retained.
In the no-hints condition, \toolname uses \code{claude-opus-4-8} to discharge (prove) \nnohintclosedB{} of \nstartreal{} proof obligations, each marked by a source-level \admits{} placeholder, including obligations left open by the reference proof, at a recorded API cost of \$\nnohintcostusd{}.
For both conditions, \code{admit.py} replaces proof bodies with \admits{} while preserving signatures, contracts, and executable code.
Success requires no new axioms or specification changes and a successful whole-crate \verus{} check; the reference measures coverage, while \verus{} establishes correctness.
\nina{it is not clear what we score against the human proof: if the agent's output already verifies, why do we need additional scoring? }
\ninadone{answered directly: the reference proof measures coverage; verification is already required by the gates.}

Prior \verus{} proof-completion evaluations score standalone function- or file-level targets with supplied specifications (\secref{sec:related}).
This run instead spans a shared production-crate dependency graph, including proof obligations inside the trusted library (the fixed, human-supplied field specifications, arithmetic facts, trusted axioms, and \code{vstd}), and accepts the run only when the final tree passes whole-crate verification.
Its coverage therefore measures integrated proof completion rather than a directly comparable per-task success rate.
Across this broad scope, the no-hints run leaves \nnohintgapsB{} particularly difficult obligations incomplete.
They lie in the deep Ristretto/Lizard curve-algebra core.
The reference leaves \ngaps{} obligations open under the same executable code and specifications, including these \nnohintgapsB{} obligations.
The final tree passes a whole-crate \verus{} check with trusted axioms and specifications unchanged.
Only gate-verified closures are credited; rejected attempts are excluded.
Most obligations close during a steady initial pass over the crate (the main sweep), where the median admit closes in \nnohintmedmin{} minutes of proving.
A later retry phase at the hard frontier spends 30--60 minutes per remaining admit but adds little, leaving the gaps (\figref{fig:nohint-dynamics}).
Here, active proving time is the sum of module-run durations, including failed attempts but excluding idle gaps between runs.
The verifier (compile, encode, and solve) accounts for {\nsolverpct}\% of active proving time and the agent, including generation and harness overhead, for {\nagentpct}\%.
We estimate time per removed admit from module-level runtimes because individual proof obligations were not timed separately.
The recorded traces also do not distinguish agent reasoning time from code-generation time.
Across its \nrunsthree{} runs at \$\ncovcostusd{} total over \ncovrounds{} rounds, the with-hints condition reached the same \nfinalgaps{}-gap residual as the no-hints run.
In these runs, retaining the doc-comment hints did not change the observed set of remaining obligations.
The shared residual obligations mark a solver frontier rather than a demonstrated impossibility: the same nonlinear field algebra remains open in the reference proof.
As an axiom-gate ablation, the campaign's prompt-only configuration (\secref{sec:campaign}) allowed \ncampfabaxioms{} fabricated axioms.
% Trimmed per Livia's 2026-07-08 directive (paper != journey): the campaign natural-ablations
% paragraph (lane-scoped plateau, contaminated prompt, dual-agent audit catches, coordinator
% ablated out) moved to the artifact record. Its two todos (one-row-per-functionality table;
% AlphaGeometry-style attribution ladder) are parked in the T29 ledger thread pending Livia.

\begin{publicationwidefigure}[t]
  \centering
  \includegraphics[width=\linewidth]{\plotasset{nohint_dynamics}}
  % Internal reminder for a later version: consider a curve in addition to the bars.
  % Internal reminder for a later version: consider adding per-area round counts, gate firings, and decomposition escalations to the artifact ledger.
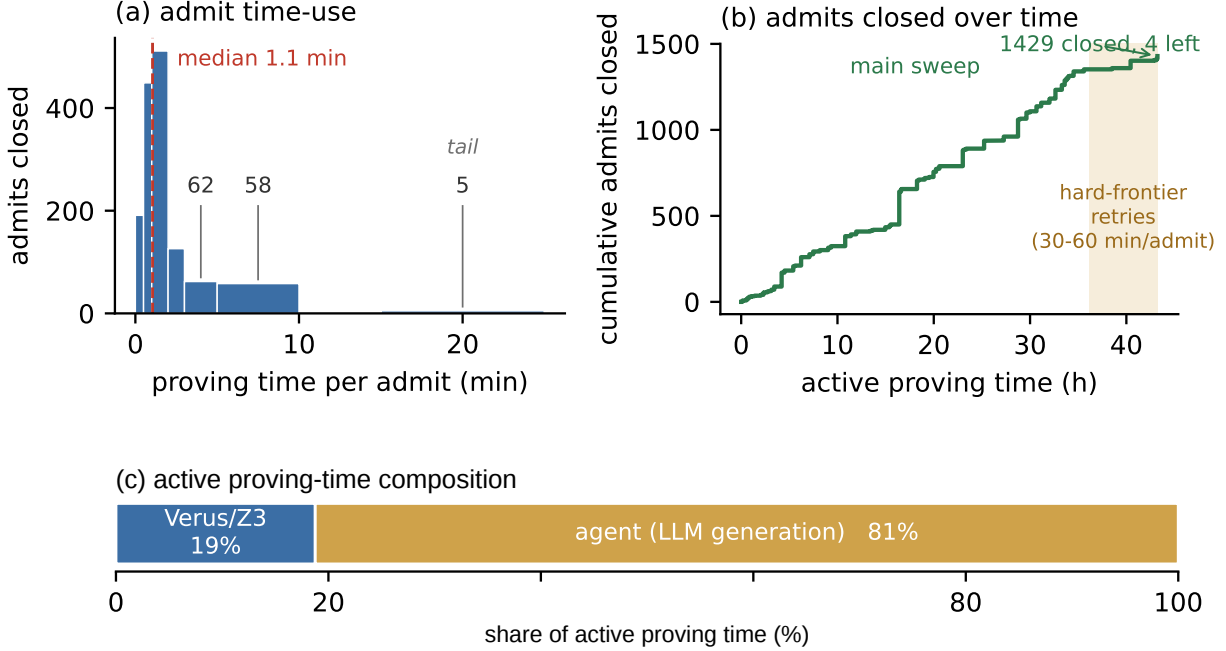
  \caption{No-hints run dynamics: (a)~minutes of proving per closed admit; (b)~cumulative admits closed against active proving time, idle and rate-limit gaps removed; (c)~the verifier's share of active proving time.
    All \nnohintclosed{} gate-verified closures are shown; rejected attempts are excluded.}
  \label{fig:nohint-dynamics}
\end{publicationwidefigure}

\FloatBarrier

\section{Proof Style: Human vs.\ Agent}
\label{app:proof-style}

% Source: codebase-comparison/style_artifacts/contract_control.{py,json}
% (blocking_drift 0, generated_lemma_contract_drift 0; re-checked 2026-07-14).
% Census re-verified 2026-07-14: 118 = |curve25519-dalek/src/**/*.rs| in the human tree,
% and all 118 are present in both agent trees (the earlier 126/121 counts included
% files outside src/, e.g. tests and benches, which the pass does not compare).
This appendix compares the proof \emph{style} of the human reference proof and the agent's no-hints run (\cref{sec:eval:coverage-sec}) under the same frozen contracts and spec-function bodies.
A tree-to-tree contract-equivalence pass compares clause text, after removing comments, across all 118 \code{.rs} files under \code{curve25519-dalek/src}, each present in both trees.
It finds zero changed \requires/\ensures{} clauses and zero redefined spec-function bodies.
The no-hints tree therefore verifies against the same frozen contracts and spec-function bodies as the fork-point human tree (the verified reference tree from which the experiment branched).
\Cref{tab:proof-style} measures how each tree constructs proofs against that shared specification, aggregated over every \code{proof fn} body and inline \code{proof \{\}} block in each tree.
\code{rlimit} raises the solver's per-query resource budget; \code{decreases} declares a termination measure.
The comment-line row measures the final artifacts; the remaining lexical metrics use a comment-stripped code view so comments and string literals do not inflate their counts.
The no-hints start tree contained no comments, but the run predates command logging, so the provenance of comments in its final artifact is not independently established.
Three patterns dominate.
The agent decomposes more finely, using more helper \code{proof fn}s in less total proof code.
On the \nstylesharedlemmas{} lemmas present in both trees, the agent version is shorter at the median: \nstylemedagentloc{} vs \nstylemedhumanloc{} proof lines.
Where the human typically justifies an assertion by calling a lemma inside that assertion's \code{assert(..) by \{ lemma() \}} block, the agent instead calls lemmas sequentially and follows them with a bare assert, using about half as many \code{by \{\}} blocks.
It also leans on solver automation where the human reasons equationally: \code{by (nonlinear\_arith)} use rises sharply, while the agent's proofs contain almost no \code{calc} blocks and far fewer \code{reveal} statements.

\begin{table}[t]
  \centering
  \small
  % Source: codebase-comparison/style_artifacts/{style_metrics.py, aggregate.py} over the
  % human / no-hints comparison trees (all proof-fn bodies + inline proof blocks,
  % counted on a comment-stripped code view; calibrated against hand counts).
  % Every cell reproduces as a column sum of style_artifacts/metrics_per_file.csv
  % (human vs agent_nh; re-verified 2026-07-14).
  \begin{tabular}{lrrr}
    \toprule
    Metric & Human & No-hints & NH/H \\
    \midrule
    \code{proof fn} count            & \npsProofFnHuman & \npsProofFnAgent & \npsProofFnRatio \\
    proof LOC                        & \npsLocHuman & \npsLocAgent & \npsLocRatio \\
    \code{assert}                    & \npsAssertHuman & \npsAssertAgent & \npsAssertRatio \\
    \code{assert .. by \{\}}         & \npsAssertByHuman & \npsAssertByAgent & \npsAssertByRatio \\
    lemma calls                      & \npsLemmaCallsHuman & \npsLemmaCallsAgent & \npsLemmaCallsRatio \\
    \code{broadcast use}             & \npsBroadcastHuman & \npsBroadcastAgent & \npsBroadcastRatio \\
    \code{forall}                    & \npsForallHuman & \npsForallAgent & \npsForallRatio \\
    \code{calc}                      & \npsCalcHuman & \npsCalcAgent & \npsCalcRatio \\
    \code{by (nonlinear\_arith)}     & \npsNonlinearHuman & \npsNonlinearAgent & \npsNonlinearRatio \\
    \code{by (bit\_vector)}          & \npsBitvectorHuman & \npsBitvectorAgent & \npsBitvectorRatio \\
    \code{reveal}                    & \npsRevealHuman & \npsRevealAgent & \npsRevealRatio \\
    \code{rlimit} attributes         & \npsRlimitHuman & \npsRlimitAgent & \npsRlimitRatio \\
    \code{decreases}                 & \npsDecreasesHuman & \npsDecreasesAgent & \npsDecreasesRatio \\
    comment lines                    & \npsCommentsHuman & \npsCommentsAgent & \npsCommentsRatio \\
    \bottomrule
  \end{tabular}
  \caption{Proof-style metrics for the fork-point human proof and the no-hints run under the same frozen contracts and spec-function bodies.
  Ratios are agent over human.}
  \label{tab:proof-style}
\end{table}
\FloatBarrier

\begin{publicationwidetable}[t]
  \centering
  \small
  \renewcommand{\arraystretch}{1.2}
\begin{adjustbox}{max width=\linewidth}
  \begin{tabular}{@{}l>{\raggedright\arraybackslash}p{5.6cm}>{\raggedright\arraybackslash}p{5.6cm}@{}}
    \toprule
    Aspect & Human reference & Agent (no-hints) \\
    \midrule
    Reasoning        & equational (\code{calc}, \code{reveal})   & solver-driven (\code{nonlinear\_arith}) \\
    Decomposition    & coarser, longer bodies                     & finer, shorter bodies \\
    Point-of-use     & \code{assert .. by \{ lemma() \}}          & sequential lemma calls, then a bare assert \\
    Stability        & small scoped steps                         & large nonlinear goals can be fragile \\
    Solver budget    & smaller local queries                      & a large query can exhaust \code{rlimit} \\
    Debuggability    & a scoped failure identifies the broken step & a resource-limit failure may not identify the step \\
    Readability      & explicit algebraic chain                   & solver call hides the algebraic steps \\
    \bottomrule
  \end{tabular}
\end{adjustbox}
  \caption{Codebase-wide proof-style contrasts (first three rows) and Montgomery-specific tradeoffs (remaining rows).}
  \label{tab:proof-style-qualitative}
\end{publicationwidetable}

\paragraph{Caveat: proof length and automation are not quality metrics.}
Lower proof LOC and heavier automation are trades, not wins.
The reference proof in \code{montgomery\_reduce\_\allowbreak part1\_\allowbreak chain\_lemmas.rs} shows why.
It uses a deliberately long multiplication ladder of small, scoped \code{assert .. by \{\}} steps and no \code{nonlinear\_arith} calls.
The no-hints proof instead calls \code{nonlinear\_arith} repeatedly.
In the reference file, the ladder reduces solver-resource demand and localizes failures.
Because nonlinear integer arithmetic is undecidable, \code{nonlinear\_arith} uses heuristic search that can exhaust the resource limit on a large goal.
The ladder instead gives the solver a sequence of local facts: distribute one limb, normalize by commutativity, and extend the prefix.
Explicit lemma calls can also supply the concrete instantiations of universally quantified facts instead of leaving the solver to search for them.
A failing scoped assert identifies the broken algebraic step; when a large nonlinear query exhausts the resource limit, its diagnostic does not identify that step.
The agent's near absence of \code{calc} chains and \code{reveal} statements also reduces readability.
An equational \code{calc} block spells out each algebraic step, whereas a bare \code{nonlinear\_arith} call leaves the reader to reconstruct why the goal holds.
The agent proofs as shipped do verify.
The codebase-wide analysis shows that the agent proofs are shorter and rely more heavily on solver automation.
The Montgomery example illustrates the resulting tradeoffs: greater solver-resource demands, less informative failures, and reasoning that is harder to follow (\tabref{tab:proof-style-qualitative}).
\FloatBarrier

\section{Parallel Orchestration Details}
\label{app:parallel}

This appendix presents the design and measurements of the parallel orchestration layer.

\subsection{Design}
\label{sec:parallel}

A proof-synthesis run over a real Rust codebase comprises hundreds of proof obligations.
The driver groups these obligations into module-level targets (\secref{sec:impl:driver}); the orchestrator schedules each target as one job.
Sequentially, the run's elapsed time is the sum of the per-job times.
A single job's time splits into two parts: \emph{agent latency} (waiting on the model) and \emph{verifier work} (\code{cargo verus} and Z3 checking the edits).
The first dominates: each job is mostly blocked on the model, while the verifier runs only in short, intermittent CPU bursts.
The workload therefore parallelizes well: while one job blocks on the model, another can use the CPU.
During local proof checking, \verus{} checks a called lemma against its signature, so bodies can be attempted in parallel (``Why Fan-Out Works'' below).
These attempts are speculative: a target can verify while lemmas it calls remain admitted, and the collected tree is accepted only after post-merge and whole-crate checks.
We therefore design an orchestration layer in \toolname with three properties.
The layer is \emph{thin}: a wrapper over the unmodified driver suffices, with no coordinator or message layer.
It is \emph{fully isolated}: naive fan-out on a single checkout is unsafe, as concurrent jobs contend on the build tree and shared state, so each worker process gets its own git worktree and results root.
It is \emph{self-balancing}: a dynamic ready-queue with longest-processing-time ordering and file decomposition keeps worker processes saturated.
The layer reaches $\nparspeedup\times$ at four-way parallelism, cutting a \nparseqmin-minute workload to \nparwallmin{} minutes.

\subsubsection{Why Fan-Out Works: Proof Bodies Can Be Attempted Independently}
\label{sec:parallel-flat}

Proof-completion targets are far more independent than their call structure suggests.
A lemma-call graph appears sequential because lemma $A$'s proof calls $B$.
During local proof checking, however, \verus{} checks $A$ against $B$'s signature, including its \ensures, even while $B$ remains \admits.
Three consequences follow.
There is no runtime propagation: each worker process proves against its own admitted copies, and proving $B$ later requires no change to $A$.
``Done'' is a global condition --- zero non-axiom \admits across the collected tree --- not a proving order.
Shared helper edits introduce merge-time coupling because two jobs may both append new helper lemmas to the same \code{lemmas/} file.
Whole-crate analyses can also reintroduce coupling, chiefly through termination checking, which requires recursive call chains to decrease a measure (``Acceptance after Recombination'' below).
Tasks that synthesize specifications introduce dependencies not present when the agent fills proof bodies against fixed specifications.
File decomposition likewise makes every worker process edit the same large file.
For a proof-completion task over a well-specified codebase with local, sparse shared helpers, these couplings are sparse enough for wide fan-out to generate candidate proofs; the post-merge whole-crate gate determines whether their combination is accepted.

\subsubsection{Scheduling and Straggler Mitigation}
\label{sec:parallel-straggler}

The orchestrator is a deterministic Python wrapper that maintains a work queue and a pool of free worker processes.
It resets each worker process before reuse and fills a free slot as soon as the next job is ready.
This dynamic queue lets a short job use a freed worker process while a long job is still running.
Even so, bounded fan-out cannot beat $\text{makespan} \ge \max_i(\text{time}_i)$: the run is no faster than its single longest job.
Scheduling determines whether that job runs concurrently with the others or starts late and extends the total elapsed time.
Two techniques reduce this straggler effect; we describe the cheaper one first.

\paragraph{Longest-processing-time (LPT) queue ordering.}
The runner sorts the queue by a difficulty proxy: descending non-axiom \admits count per job.
This ordering dispatches known-heavy jobs in the first wave, so they run concurrently with the other jobs rather than finishing after the other worker processes become idle.
This is LPT list scheduling ($\approx\!4/3$-optimal for makespan)~\cite{graham1969bounds} and costs only a sort.

\paragraph{File decomposition.}
When a \emph{single file} dominates the makespan, LPT ordering cannot help because the file is one job.
The signature argument above still makes its proof bodies independently attemptable.
Each \admits-bearing \code{proof fn} can verify against the other functions' admitted signatures in a separate worktree before recombination.
The runner LPT-packs the functions, weighted by admit count, into \(K\) groups, approximately the number of worker processes.
It proves each group in the \emph{same} admitted file and scopes the completion gate to the assigned functions through the driver's \code{--only-fns} flag.
The remaining functions stay admitted and supply their \ensures.
The runner then splices each proven body back into one file and unions the helper lemmas introduced by every group.
It three-way merges only the small additive import header because a whole-file merge would scatter conflicting hunks.
The merged file is then re-verified once with a module-scoped \verus{} gate (``Acceptance after Recombination'' below).

\subsubsection{Acceptance after Recombination}
\label{sec:parallel-limits}

A file is reported proved only if \verus{} verifies it after recombination.
The independence argument of ``Why Fan-Out Works'' holds for types and specifications but not for \verus{}'s whole-crate analyses.
In particular, lemmas that verify in isolation can form a recursion cycle after their bodies are spliced together, causing the combined tree to fail termination checking.
A per-group ``complete'' label is therefore not evidence that the whole file is proved.
The runner therefore accepts a decomposition only when a post-merge, module-scoped \verus{} check re-verifies the entire recombined file on a clean worktree with zero non-axiom admits.
\secref{sec:parallel-results} reports one recombined file this check rejected.
The module-scoped check is a filter, not the final authority: acceptance of the collected tree still requires the whole-crate \verus{} check, which alone covers crate-level analyses.
Strongly-connected-component-aware grouping, sourcing each group's last verifier-clean round, and carrying newly introduced helper lemmas across the splice improve the chance that recombination succeeds on the first attempt.

\subsection{Measured Speedup}
\label{sec:parallel-results}
We measured speedup across four configurations, each adding scale or machinery to the last and each measured once.
Because the workload is API-bound, the relevant figure is the \emph{like-for-like} speedup: the sum of the participating jobs' own durations (the sequential cost of the same work) divided by the parallel makespan.
\tabref{tab:parallel-speedup} summarizes the measurements.

\begin{publicationwidetable}[t]
  \centering
  \caption{Like-for-like speedup across orchestration configurations, each measured once.
  The speedup ceiling is $N$.
  Speedups are computed from unrounded durations, so the last digit can differ from the quotient of the rounded columns.}
  \label{tab:parallel-speedup}
  \small
  \setlength{\tabcolsep}{4pt}
  % Source: data/appendix_results.json, generated into results-macros.tex.
\begin{adjustbox}{max width=\linewidth}
  \begin{tabular}{@{}llrrrrr@{}}
    \toprule
    Configuration & Scheduling & $N$ & Jobs & Seq.\ sum & Makespan & Speedup \\
    \midrule
    Two identical synthetic jobs & static assignment    & \nparTwinsWorkers & \nparTwinsJobs & \nparTwinsSeqSecs\,s & \nparTwinsWallSecs\,s & $\nparTwinsSpeedup\times$ \\
    Synthetic pool test         & dynamic ready-queue  & \nparPoolWorkers & \nparPoolJobs & \nparPoolSeqSecs\,s & \nparPoolWallSecs\,s & $\nparPoolSpeedup\times$ \\
    Eight real modules          & dynamic, per-round   & \nparEightWorkers & \nparEightJobs & \nparEightSeqMin\,min & \nparEightWallMin\,min & $\nparEightSpeedup\times$ \\
    Decomposed file $+$ mixed (real) & LPT + decomposition  & \nparMixedWorkers & \nparMixedJobs & \nparseqmin\,min & \nparwallmin\,min & $\mathbf{\nparspeedup\times}$ \\
    \bottomrule
  \end{tabular}
\end{adjustbox}
\end{publicationwidetable}

\paragraph{Observed speedup and scheduling effects.}
No configuration triggered API rate-limiting.
The two-identical-job configuration reached $\nparTwinsSpeedup\times$; the jobs' per-round times (\nparTwinsFastSecs/\nparTwinsSlowSecs\,s) bracketed the solo time (\nparTwinsSoloSecs\,s).
The sub-$2\times$ result is consistent with a straggler from model nondeterminism, but one measurement cannot separate that explanation from other causes.
The dynamic ready-queue reached $\nparPoolSpeedup\times$, and a freed slot refilled within \nparPoolRefillSecs\,s via reset-before-reuse.
In the eight-module measurement, a plain queue reached $\nparEightSpeedup\times$; the heaviest module (\nparHeavyAdmits{} admits) sat last in the input and formed the observed tail.
Adding LPT ordering and splitting that heavy file into three parallel groups reached $\nparspeedup\times$ on the final configuration, cutting \nparseqmin{} minutes to \nparwallmin{}.
The remaining gap to $\nparCeiling\times$ is consistent with load imbalance and decomposition overhead, but the single measurement does not isolate their contributions.

\paragraph{One decomposed file failed after recombination.}
Splitting \code{batch\_compress\_lemmas} (\nparHardTotal{} lemmas) \nparHardGroups{} ways proved \nparHardProved{} of \nparHardTotal{} independently.
The merged file nevertheless failed the post-merge \verus gate.
The spliced-in bodies formed a recursion cycle lacking the \code{decreases} clause required by the combined call graph (the termination coupling described under ``Acceptance after Recombination'', \secref{sec:parallel}).
The gate rejected the recombined file rather than reporting a false success.
Independently verified groups therefore do not establish termination of the combined tree.
The recombined file must pass its post-merge module-scoped check.
Final acceptance still requires the whole-crate check.

\FloatBarrier

\section{Module Placement and Manifest Detail}
\label{app:placement}

\begin{publicationwidefigure}[t]
\centering
\definecolor{ffCodeFill}{HTML}{E5E7EB}
\definecolor{ffCodeDraw}{HTML}{6B7280}
\definecolor{ffSpecFill}{HTML}{D9F0FB}
\definecolor{ffSpecDraw}{HTML}{0072B2}
\definecolor{ffProofFill}{HTML}{D8F3EA}
\definecolor{ffProofDraw}{HTML}{009E73}
\definecolor{ffGateDraw}{HTML}{9A4F86}
\definecolor{ffFloorDraw}{HTML}{E69F00}
\begin{adjustbox}{max width=\linewidth}
\begin{tikzpicture}[
  font=\footnotesize,
  flow/.style={-{Stealth[length=2.1mm]}, semithick},
  sheet/.style={rounded corners=1pt, line width=0.35pt},
  codeSheet/.style={sheet, draw=ffCodeDraw, fill=ffCodeFill},
  specSheet/.style={sheet, draw=ffSpecDraw, fill=ffSpecFill},
  proofSheet/.style={sheet, draw=ffProofDraw, fill=ffProofFill},
  missingProof/.style={sheet, draw=ffProofDraw, dashed, fill=white},
]
  \def\w{4.60}
  \def\rowH{0.82}
  \def\xA{0.00}
  \def\xB{8.20}
  \def\xMid{6.40}

  \node[font=\bfseries] at (\xA+2.30,4.70) {start state};
  \node[text=ffProofDraw] at (\xA+2.30,4.44) {lemma contracts + proofs stripped above the trusted library};
  \node[font=\bfseries] at (\xB+2.30,4.70) {audited end state};
  \node[text=ffProofDraw] at (\xB+2.30,4.44) {lemma contracts + proofs regenerated};

  % above the floor: code frozen, spec + proof stripped (start) / regenerated (end)
  \foreach \name/\y in {public APIs/3.00, curve + scalar/2.00} {
    \draw[draw=black!35, fill=black!2, rounded corners=1pt] (\xA,\y) rectangle (\xA+\w,\y+\rowH);
    \node[anchor=east] at (\xA-0.14,\y+0.42) {\name};
    \draw[codeSheet]    (\xA+0.18,\y+0.08) rectangle (\xA+\w-0.46,\y+0.42);
    \draw[specSheet]    (\xA+0.30,\y+0.21) rectangle (\xA+\w-0.34,\y+0.55);
    \draw[missingProof] (\xA+0.42,\y+0.34) rectangle (\xA+\w-0.22,\y+0.68);
    \node[text=ffProofDraw, font=\scriptsize, fill=white, inner sep=1pt] at (\xA+2.40,\y+0.51) {\admits};
    \draw[draw=black!35, fill=black!2, rounded corners=1pt] (\xB,\y) rectangle (\xB+\w,\y+\rowH);
    \draw[codeSheet]  (\xB+0.18,\y+0.08) rectangle (\xB+\w-0.46,\y+0.42);
    \draw[specSheet]  (\xB+0.30,\y+0.21) rectangle (\xB+\w-0.34,\y+0.55);
    \draw[proofSheet] (\xB+0.42,\y+0.34) rectangle (\xB+\w-0.22,\y+0.68);
  }

  % trusted floor (the orange line) and everything below it: frozen substrate
  \draw[ffFloorDraw, semithick] (\xA-0.02,1.91) -- (\xA+\w,1.91);
  \draw[ffFloorDraw, semithick] (\xB-0.02,1.91) -- (\xB+\w,1.91);
  \node[anchor=east, text=ffFloorDraw, font=\scriptsize, fill=white, inner sep=1pt] at (\xA-0.14,1.91) {trusted library};

  \foreach \name/\y in {field + common/1.00, backend/0.00} {
    \draw[draw=black!35, fill=black!2, rounded corners=1pt] (\xA,\y) rectangle (\xA+\w,\y+\rowH);
    \node[anchor=east] at (\xA-0.14,\y+0.42) {\name};
    \draw[codeSheet]  (\xA+0.18,\y+0.08) rectangle (\xA+\w-0.46,\y+0.42);
    \draw[specSheet]  (\xA+0.30,\y+0.21) rectangle (\xA+\w-0.34,\y+0.55);
    \draw[proofSheet] (\xA+0.42,\y+0.34) rectangle (\xA+\w-0.22,\y+0.68);
    \draw[draw=black!35, fill=black!2, rounded corners=1pt] (\xB,\y) rectangle (\xB+\w,\y+\rowH);
    \draw[codeSheet]  (\xB+0.18,\y+0.08) rectangle (\xB+\w-0.46,\y+0.42);
    \draw[specSheet]  (\xB+0.30,\y+0.21) rectangle (\xB+\w-0.34,\y+0.55);
    \draw[proofSheet] (\xB+0.42,\y+0.34) rectangle (\xB+\w-0.22,\y+0.68);
  }

  \draw[specSheet] (\xA+0.42,3.70) rectangle (\xA+\w-0.22,3.82);
  \draw[specSheet] (\xB+0.42,3.70) rectangle (\xB+\w-0.22,3.82);
  \node[anchor=east, text=ffSpecDraw, font=\scriptsize, fill=white, inner sep=1pt] at (\xA+\w-0.26,4.12) {API contract};
  \node[anchor=east, text=ffSpecDraw, font=\scriptsize, fill=white, inner sep=1pt] at (\xB+\w-0.26,4.12) {API contract};

  \draw[ffGateDraw, dashed, semithick] (\xA+0.10,-0.10) rectangle (\xA+\w-0.10,3.88);
  \draw[ffGateDraw, dashed, semithick] (\xB+0.10,-0.10) rectangle (\xB+\w-0.10,3.88);
  \node[text=ffGateDraw, align=center] at (\xA+2.30,-0.40) {frozen code, contracts + trusted library};
  \node[text=ffGateDraw, align=center] at (\xB+2.30,-0.40) {same frozen surface};

  \draw[flow, ffProofDraw] (\xA+\w+0.30,2.55) -- (\xB-0.30,2.55);
  \node[align=center, text=ffProofDraw, font=\scriptsize, fill=white, inner sep=1.5pt] at (\xMid,2.92)
    {\toolname{}\\writes lemma contracts + proofs};
  \draw[flow, ffGateDraw] (\xB-0.30,1.25) -- (\xA+\w+0.30,1.25);
  \node[align=center, text=ffGateDraw, font=\scriptsize, fill=white, inner sep=1.5pt] at (\xMid,0.88)
    {\verus{} + gates\\check frozen surface};
\end{tikzpicture}
\end{adjustbox}
\caption{Start and audited end states of the proof-and-spec synthesis run.}
\label{fig:field-floor-mode}
\end{publicationwidefigure}

This appendix expands \figref{fig:peel-design}'s given/synthesized summary into a per-module placement map and the detail of each \emph{manifest}.
A manifest is the per-experiment file that defines the experiment's cut: the specification and proof material removed for regeneration.
For each module, the manifest records whether the transform keeps it editable, deletes its lemmas, strips its proof bodies, or freezes it.
\figref{fig:field-floor-mode} shows the start and audited end states of the proof-and-spec synthesis run at that granularity.
The audit accepts only a whole-crate \verus{} success whose gates confirm that the fixed inputs shown in \figref{fig:field-floor-mode} are unchanged.

\paragraph{Module placement.}
\yzj{This paragraph is too long; break it into several shorter ones --- e.g.\ one for the above-floor rows, one for the floor rows, one for the cross-cutting specification/contract material, and one for the operation-level distinction and floor definition.}\resolve{claude: split into the four groups yzj proposed, text unchanged; approved by Livia 2026-07-28.}
The rows of \figref{fig:field-floor-mode} group modules by the material a manifest can remove from them; they are not a complete module hierarchy.\yzj{``there'' $\to$ ``from them''.}\resolve{claude: adopted yzj's referent fix; approved under Livia's 2026-07-28 delegation.}
The public-API row contains caller-facing modules whose public contracts are the fixed boundary: \code{edwards.rs}, \code{montgomery.rs}, \code{ristretto.rs}, and \code{scalar.rs}.
API-adjacent glue such as \code{traits.rs} and \code{window.rs} lives on the same public-facing surface when a manifest strips proof bodies from those files.
The curve/scalar row contains the helper-lemma material for the modules above the trusted library: the edwards, ristretto, scalar, and scalar-byte lemma directories under \code{lemmas/**}.
It also includes multiscalar and scalar-multiplication proof code, plus the curve-model, Montgomery, Jacobi-quartic, Ristretto, and Lizard obligations that consume those facts.

The field/common row contains the field arithmetic proof layer and reusable arithmetic substrate: \code{lemmas/field\_lemmas/**} and \code{lemmas/common\_lemmas/**} (the number-theory, pow, div--mod, mask, bit, shift, multiplication, sum, and \code{to\_nat} helper lemmas).
The backend/trusted row contains the backend field/scalar implementations, \code{vstd}, and the trusted \axiomfn{} lemmas.

Specification material crosses all rows: \code{specs/**} and spec functions embedded in API or lemma modules are spec material, while public \requires/\ensures clauses on caller-facing functions are contract material.

Executable Rust code is frozen in every experiment discussed here; when a manifest lists an executable module with \code{strip-all}, the transform removes proof-only bodies inside that module, never executable bodies.
Removing a proof body and deleting a helper lemma together with its statement differ only in which text the transform removes.
The distinction is operational, not a conceptual layering of the crate.
The trusted library is the fixed collection defined in \secref{sec:bg:layers}: field specifications, field/common arithmetic facts, trusted \axiomfn{} lemmas, and \code{vstd}.
A manifest chooses the editable region outside the trusted library; the common-arithmetic files are part of that library in every run discussed here, and a field-arithmetic repair run could instead make the field layer itself the target.

% \paragraph{Manifest record.}
% The supplementary artifact's proof-and-spec synthesis manifest records the exact per-file placement of the \nfieldfloorfiles{}-file editable set; it also records the placement of the frozen Lizard modules, the seven top-level Montgomery/scalar lemma files, and all \code{mod.rs} files.

\paragraph{Scope comparison.}
The whole-crate proof-only run of \secref{sec:eval:setup} removes proof bodies crate-wide and nothing else.
The proof-and-spec synthesis run of \secref{sec:noapi} removes proof bodies, helper lemmas, and internal specifications from the Edwards/Montgomery/Ristretto/scalar region named above, while contracts and the trusted library stay fixed.

\FloatBarrier
\section{Proof-and-Spec Synthesis Family: Run Detail}
\label{app:certdetail}

This appendix carries the run-level detail behind \secref{sec:noapi}: the run protocol, verifier use, run dynamics and repairs, and the recorded integrity events.
The \emph{convergence ladder} split the proof-and-spec synthesis run's editable region into per-file targets and accepted each target only after its audit, before final whole-crate acceptance.
The artifact's ladder record calls an accepted target a \emph{bank} and the final acceptance the \emph{seal}.\yzj{Not sure how these three items map to the $\sim$11 paragraphs below, which cover more than the roadmap names. Either expand this roadmap to match, or group the paragraphs under the three promised headings so the reader can follow the structure.}\resolve{claude: expanded the roadmap to name the section's actual four groups; the run-family table was removed in T240; approved by Livia 2026-07-28.}\resolve{claude: added the four subsection headings yzj requested, matching the roadmap; approved by Livia 2026-08-01.}

In an earlier, smaller pilot that left both API files editable, the post-run audit found the API contracts byte-identical to the reference; the \specdrift{} and \gitrecovery{} gates, not agent restraint, held that boundary.

\subsection{Run Protocol}
\paragraph{Prompt and feedback.}
The fixed prompt included general guidance about ordering, contract construction, decomposition, and scale; the decomposition guidance reflected the codebase's proof style.
The prompt contained no proof content, lemma names, inventories, or target-specific ordering.
Per-round feedback contained only verifier errors and the agent's current admit inventory.
The editable files nevertheless retained roughly \ncertcommentlines{} lines of comments inherited from the verified source.
A start-state content audit found mostly module headers, section dividers, and specification documentation of the encoding layout, plus three proof-strategy comments inherited from the human reference.
% Source: dalek-lite-mvp/docs/run_stats/stage3_certificate_record.md (pre-registered attempts
% table; run ids *_480_*; attempt 1 = clean budget-exhaustion LIMIT, zero gate events;
% "No hint escalation was ever used: the pre-registered H1--H4 ladder went untouched").
Stopping rules and the success predicate were the driver's (\cref{lst:driver}).
The pre-registered protocol allowed multiple attempts with a 480-minute wall-clock budget per attempt.

\subsection{Verifier Use}
\paragraph{Verifier utilization.}
Colored bars in \figref{fig:wallclock-share} show active elapsed time with at least one whole \code{cargo verus} call in flight; gray shows the remaining agent and orchestration time.
Because \toolname{} runs verifier calls sequentially, its colored share also equals total verifier-call time divided by elapsed time.
Claude Code alone launched verifier calls from the main thread and subagents, sometimes concurrently.
The bar counts overlapping calls once, while the annotation adds the duration of every call and reports the total in verifier-hours.
Inferred start times make the Claude Code measurements slight upper bounds.
These descriptive measurements do not explain completion or isolate the effect of any driver, skill, or gate.

\begin{publicationwidefigure}[t]
  \centering
  % Data: data/appendix_results.json, generated into results-macros.tex.
  \definecolor{wcCP}{HTML}{009E73}
  \definecolor{wcCC}{HTML}{E69F00}
  \definecolor{wcBg}{HTML}{E5E7EB}
\begin{adjustbox}{max width=\linewidth}
  \begin{tikzpicture}[font=\footnotesize]
    \def\xs{0.075}   % cm per percentage point (100 -> 7.5cm)
    \def\bh{0.5}     % bar height
    % --- background + colored bars ---
    \fill[wcBg] (0,1.2) rectangle (100*\xs,1.2+\bh);
    \fill[wcCP] (0,1.2) rectangle (\ncertsolverpct*\xs,1.2+\bh);
    \fill[wcBg] (0,0.2) rectangle (100*\xs,0.2+\bh);
    \fill[wcCC] (0,0.2) rectangle (\nbasewallcovpct*\xs,0.2+\bh);
    % --- row labels ---
    \node[anchor=east] at (-0.15,1.45) {\toolname};
    \node[anchor=east] at (-0.15,0.45) {\claudecode{} alone};
    % --- inline annotations ---
    \node[anchor=west, xshift=2pt] at (\ncertsolverpct*\xs,1.45) {\ncertsolverpct\% solver share; agent + orchestration \ncertagentpct\%};
    \node[anchor=west, xshift=2pt, align=left] at (\nbasewallcovpct*\xs,0.45) {\nbasewallcovpct\%;\\ \nbaseverushours{} summed verifier-hours (overlapping)};
    % --- x-axis ---
    \draw (0,0) -- (100*\xs,0);
    \foreach \p in {0,20,40,60,80,100}{%
      \draw (\p*\xs,0) -- (\p*\xs,-0.08);
      \node[anchor=north, font=\scriptsize] at (\p*\xs,-0.1) {\p};
    }
    \node[anchor=north] at (50*\xs,-0.5) {share of elapsed time spent in the verifier};
  \end{tikzpicture}
\end{adjustbox}
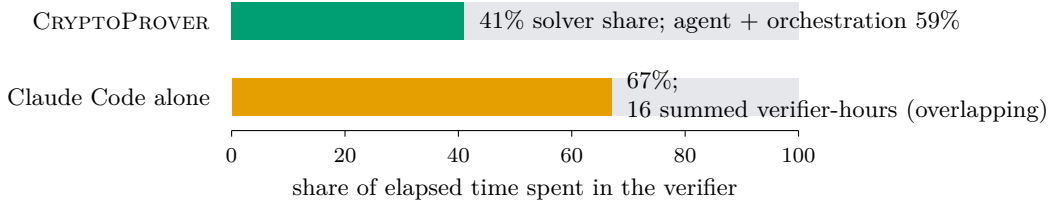
  \caption{Share of elapsed time spent in the verifier for the proof-and-spec synthesis comparison.}
  \label{fig:wallclock-share}
\end{publicationwidefigure}

\paragraph{Solver attributes at acceptance.}
The proof-and-spec synthesis run's final tree contains solver limits larger than the largest limit used by the human reference.
At final whole-crate acceptance, targeted single-attribute-removal checks identified one raised limit that could not be removed: \code{scalar::}\allowbreak\code{non\_adjacent\_form} at \code{rlimit(150)}.
Every other limit above the human reference's maximum was individually droppable under its module check.
The \code{claude-opus-4-8} replication completed the same task in \resOpusFourEightArmDurationHours{} hours of elapsed time at \$\resOpusFourEightArmCostUSD{} in recorded API cost, versus \ncerthours{} hours for the \code{claude-fable-5} run.
In that replication, \resOpusFourEightArmRlimitLoadBearingSites{} of \resOpusFourEightArmRlimitSites{} solver-limit sites remained necessary when each raised attribute was removed individually and the affected module was re-verified.
The generated Montgomery proofs also call the pre-existing trusted-library helper \code{lemma\_u128\_shl\_is\_mul}, whose \code{assume(false)} body is documented as pending \code{vstd} support.
The run introduced no new assumptions, and the reference proof calls the same helper.
% Checked 2026-07-14: the ladder/proof-and-spec synthesis audits ran at the DEFAULT rlimit
% (ladder_stage1_results.md: "never --rlimit 80"; stage3_certificate_record.md battery item 2),
% while rlimit 80 was the earlier spec-gen runbook audit protocol (spec_gen_runbook.md).
% The opus replication's load-bearing site NAMES live only
% in the VM1 rlimit_gates_a9/*.json verdicts (see the opus block in results-macros.tex);
% name them here when that artifact is staged.

\subsection{Run Dynamics and Repairs}
\paragraph{Agent-generated internal-specification repair.}
Under an early policy that froze agent-generated internal specifications between rounds, the continuation run \code{corefloor\_006}, which resumed an earlier attempt from its saved state, stopped with \hfCoreflZeroZeroSixAdmits{} non-axiom \admits{} open.
At least two obligations were unprovable because the agent had generated false internal specifications.
The machine-checked counterexample $x=0$, $y=p+1$ falsifies one Ristretto statement: the input violates its postcondition because the precondition lacks a canonicality requirement.
The revised policy allowed the agent to correct agent-generated internal specifications while executable code, API contracts, and the trusted library remained fixed.
For \code{lemma\_carry8\_bound}, the counterexample \code{carry8} $= 2^{53}+13$ led the agent to replace the too-weak precondition \code{l4} $< 2^{52}$ with the call-site fact \code{l4} $= 2^{44}$ and complete the proof.

\paragraph{Frozen-caller references supplied the early proof order.}
\figref{fig:certificate-dynamics} traces the run's descent from the first measured whole-crate state to final whole-crate acceptance across both attempts.
The evolution record shows two phases.
A scaffold phase first declared the missing lemmas that frozen callers reference, each with its proof deferred as a \emph{scaffold admit}.
The first hour's edit order followed the concentration of frozen-caller references, supplying a dependency order absent from the generic prompt.
A bottom-up discharge phase then discharged the scaffold admits to zero.
The verifier accepted the agent's weakened version of a too-strong generated contract and its explicit witnesses for solver repairs.
The agent also inserted an exploratory ghost-construction probe and removed it before final whole-crate acceptance.

\paragraph{Decomposition closed the target after higher solver limits timed out.}
The agent first tried a monolithic step lemma at \code{rlimit(600)} and then at \code{rlimit(900)}; two checks in a row hit the verifier's wall-clock ceiling without returning a verdict.
It then extracted a closed-form helper lemma, split the two offending sublemmas, and reduced the surviving budget attributes to \code{rlimit(300)}.
The split lemmas verified within the round.
This was a structural decomposition of the kind the reference also uses, reached with no reference proof visible.

\paragraph{The run encountered four classes of solver goals.}
\verus{} calls these logical goals \emph{verification conditions}.
Type-invariant construction was the only class that prompted an exploratory proof attempt.
Loop invariants closed structurally without budget attributes.
Preconditions about numeric bounds were numerous but mechanically discharged by weakening each intermediate bound and summing the results.
Termination produced no failures in either the convergence-ladder or proof-and-spec synthesis run.

\paragraph{A verifier crash understated the remaining errors.}
During a later resume of the same continuation, \linebreak[1]\code{peel\_\allowbreak corefloor\_\allowbreak 006\_\allowbreak resume14}, a well-typed \code{array\_view} call panicked under the pinned \verus{} release in the interpreter for \verus{}'s intermediate representation (VIR) when its array remained symbolic.
The reproduced cases covered both symbolic parameters and symbolic field projections.
The panic terminated verification before the final summary, so preliminary diagnostics could not establish the remaining error count.
The \code{verus\_check} tool now marks missing-summary runs as truncated, sets the authoritative error count to unknown, and excludes those runs from plateau decisions.
A narrow repair to the pinned verifier implements the interpreter's documented fallback: it preserves a simplified residual \code{array\_view} call for symbolic inputs while leaving concrete-array reduction unchanged.
The patched verifier passed a validation run over the campaign's fully verified crate; this validation was separate from the acceptances reported in this paper, which used the unmodified pinned release.
Subsequent campaign stability is not a direct regression result for this defect: the observed proofs used scalarized values or a concrete local array, not a symbolic array parameter or field projection.
Unsupported symbolic operations can still terminate during elaboration before the final summary.
The repair therefore removes this internal panic but does not make preliminary diagnostic counts complete, and the harness continues to treat missing-summary runs as indeterminate.
The observed failures aborted before acceptance.
They exposed a verifier crash and incomplete diagnostics but did not produce an unsound accepted tree.

\subsection{Recorded Integrity Events}
\paragraph{The convergence ladder's audit record.}
\tabref{tab:verification-stack} lists each false-success threat, its countermeasure, and the outcome observed in the recorded runs.
Every catch occurred before any per-file acceptance was recorded, and all \nladderbanks{} accepted targets then survived independent fresh-container re-verification, with zero false successes and zero retractions.
For example, \frozenedit{} rejected a whole-crate verification success after the agent edited a frozen backend witness; the harness reverted the edit, and the same agent produced a tree that passed the final check in the next round.
Separately, a preflight audit traced a \code{LIMIT} to stale text in one target's rendered prompt; after the prompt defect was corrected, that target was accepted.

\begin{publicationwidetable}[t]
  \centering
  \footnotesize
  \setlength{\tabcolsep}{3.5pt}
  % Instances: dalek-lite-mvp/docs/run_stats/ladder_stage1_results.md
  % ("False-green taxonomy caught in the wild" + headline findings 2/4/5).
\begin{adjustbox}{max width=\linewidth}
  \begin{tabular}{ll>{\raggedright\arraybackslash}p{3.6cm}}
    \toprule
    False-success threat & Countermeasure & Recorded outcome \\
    \midrule
    \code{COMPLETE} echoed in the agent stream & score only the harness result & echoed claims never accepted \\
    module verifies but the crate is unchecked & whole-crate verify at default rlimit & mid-run module successes refused \\
    a compile failure hides the true error count & settled final state, not series minima & transient two-error states refused \\
    zero errors with admits still open & crate-wide non-axiom admits $=0$ & \nscaffolddrafts{} scaffold admits held to discharge \\
    edit to a frozen witness file & \frozenedit{} gate & one contaminated success reverted and re-accepted \\
    stale text in a rendered prompt & rendered-prompt preflight audit & one \code{LIMIT} traced, fixed, and accepted \\
    \bottomrule
  \end{tabular}
\end{adjustbox}
  \caption{False-success threats, countermeasures, and recorded outcomes.}
  \label{tab:verification-stack}
\end{publicationwidetable}
\FloatBarrier

% Restored by Livia's ruling 2026-08-01 (reversing T272 item 6) in the plain-language form she reviewed.
\paragraph{The gates rejected two prohibited edits in early setups.}
In the early setups that never finished the task, the gates rejected a weakened specification and an edited verifier.
The records of the later, successful runs contain no corresponding gate events.
Other conditions also changed between these setups, so the contrast is observational.\yzj{This paragraph mixes supporting evidence with a discussion-level conclusion; the conclusion (and the ``natural experiment'' framing) may belong in the main-text Discussion, leaving only the evidence here.}\resolve{claude: removed the discussion-level clause; the factual contrast stays with its evidence; approved by Livia 2026-07-28.}\resolve{claude: restored in plain language by Livia's 2026-08-01 ruling after her T272 removal.}

\end{document}